\documentclass[12pt]{iopart}
\usepackage{iopams} 
\usepackage{graphics}
\usepackage{pennames}
\usepackage{amssymb}
\usepackage{setstack}
\begin{document}
\title[Expansion dynamics in Pb-Pb collisions]{Centrality dependence of the expansion dynamics in Pb-Pb collisions at 
158~A~GeV/$c$}
%
%
%
\author{
The WA97 Collaboration:\\
F~Antinori$^{9}$, 
W~Beusch$^{5}$, 
I~J~Bloodworth$^{4}$, 
G~E~Bruno$^{1}$, 
R~Caliandro$^{1}$, 
N~Carrer$^{9}$, 
D~Di~Bari$^{1}$, 
S~Di~Liberto$^{11}$, 
D~Elia$^{1}$, 
D~Evans$^{4}$, 
K~Fanebust$^{2}$, 
F~Fayazzadeh$^{8}$, 
R~A~Fini$^{1}$, 
B~Ghidini$^{1}$,  
G~Grella$^{12}$, 
H~Helstrup$^{3}$,  
M~Henriquez$^{8}$, 
A~K~Holme$^{8}$, 
A~Jacholkowski$^{1}$, 
G~T~Jones$^{4}$, 
J~B~Kinson$^{4}$, 
K~Knudson$^{5}$, 
I~Kr\'alik$^{6}$, 
V~Lenti$^{1}$, 
R~Lietava$^{6}$, 
R~A~Loconsole$^{1}$, 
G~L\o vh\o iden$^{8}$, 
V~Manzari$^{1}$, 
M~A~Mazzoni$^{11}$, 
F~Meddi$^{11}$, 
A~Michalon$^{13}$, 
M~E~Michalon-Mentzer$^{13}$, 
M~Morando$^{9}$, 
P~I~Norman$^{4}$, 
B~Pastir\v c\'ak$^{6}$, 
E~Quercigh$^{5}$, 
D~R\"{o}hrich$^{2}$, 
G~Romano$^{12}$, 
K~\v{S}afa\v r\'{\i}k$^{5}$, 
L~\v S\'andor$^{5,6}$, 
G~Segato$^{9}$, 
P~Staroba$^{10}$, 
M~Thompson$^{4}$, 
J~Urb\'{a}n$^{6}$, 
T~Vik$^{8}$, 
O~Villalobos~Baillie$^{4}$, 
T~Virgili$^{12}$, 
M~F~Votruba$^{4}$  and P~Z\'{a}vada$^{10}$.}
\address{$^1$\  
 Dipartimento I.A. di Fisica dell'Universit\`{a} e del Politecnico
 di Bari and Sezione  INFN, Bari, Italy}
\address{$^2$\
 Fysisk institutt, Universitetet i Bergen, Bergen, Norway}
\address{$^3$\
 H\o gskolen i Bergen, Bergen, Norway} 
\address{$^4$\
 School of Physics and Astronomy, University of Birmingham, Birmingham, UK}
\address{$^5$\
 CERN, European Laboratory for Particle Physics, Geneva, Switzerland}
\address{$^6$\
 Institute of Experimental Physics, Slovak Academy of Sciences, Ko\v{s}ice, Slovakia}
\address{$^7$\
 GRPHE, Universit\'{e} de Haute Alsace, Mulhouse, France}
\address{$^8$\
Fysisk institutt, Universitetet i Oslo, Oslo, Norway}
\address{$^9$\
Dipartimento di Fisica dell'Universit\`{a} and Sezione INFN, Padua, Italy}
\address{$^{10}$\
Institute of Physics, Academy of Sciences of Czech Republic, Prague, Czech Republic}
\address{$^{11}$\
Dipartimento di Fisica dell'Universit\`{a} ``La Sapienza'' and Sezione INFN, Rome, Italy}
\address{$^{12}$\
Dipartimento di Scienze Fisiche ``E.R. Caianiello'' dell'Universit\`{a} and INFN, Salerno, Italy}
\address{$^{13}$\
Institut de Recherches Subatomiques, IN2P3/ULP, Strasbourg, France 
}
\begin{abstract}
Two-particle correlation functions of negatively charged hadrons from 
Pb-Pb collisions at 158~GeV/{\em c} per nucleon have been measured by the WA97 
experiment at the CERN SPS. A Coulomb correction procedure 
that assumes an expanding source 
has been implemented. 
Within the framework of an expanding thermalized source model the size 
and dynamical state of the collision fireball at freeze-out have been 
reconstructed as a function of the centrality of the collision. 
Less central collisions exhibit a different dynamics than central ones: 
both transverse and longitudinal expansion velocities are slower, 
the expansion duration is shorter and the system 
freezes out showing smaller dimensions and higher temperature. 
\end{abstract}
\submitto{JPG}
\maketitle
%
\section{Introduction}
\label{intro}
\par
The study of ultrarelativistic heavy ion collisions is motivated mainly 
by the QCD predictions that at sufficiently high energy density the excited 
nuclear matter undergoes a phase transition into a system of deconfined 
quarks and gluons (Quark--Gluon Plasma: QGP).
\par
The most direct measurement of space-time characteristics of the 
collision region is provided by 
Hanbury-Brown and Twiss (HBT) interferometry, 
a method developed originally in astronomy in the fifties~\cite{HBT} 
and applied independently in the field of particle physics a few years 
later~\cite{GGLP}. 
\par
Over the last years this method has been widely used in 
ultrarelativistic heavy collisions to investigate the space-time 
geometry of the source,  
and ample evidence accumulated showing that information can be gained 
both on its  geometry and its dynamics. For a recent review see 
e.g.~\cite{review}.
\par
In ref.~\cite{RecProgr,RecProgr2} an analysis 
strategy for reconstructing the freeze-out source state in heavy-ion 
collisions was suggested. A model for an 
emitting source characterized by 
a collective flow dynamics superimposed 
to a random (thermal) motion was assumed therein. 
The parameters of the model are the freeze-out temperature $T$, 
the transverse geometric (Gaussian) radius $\sqrt{2}R_G$, the 
average freeze-out proper time $\tau_0$, the mean proper time 
duration of particle emission $\Delta\tau$, the transverse 
flow velocity $\beta_{\perp}$, and $\Delta\eta$, that parametrizes 
the longitudinal extension of the source.\\
This approach has been followed by the NA49 experiment, that has 
reconstructed the freeze-out state for very central Pb-Pb collisions 
(about $3\%$\ of the most central events) at $158\, {\rm GeV}$\ per 
nucleon~\cite{NA49}.
\par
In this paper we also follow the model of 
ref.~\cite{RecProgr,RecProgr2}, 
but in addition we evaluate all the source parameters as a function 
of centrality, owing to the relatively large centrality window 
explored by the WA97 experiment.  
\par
The correlation analysis presented 
here is based on pairs of negative hadrons 
($h^-$), which are dominated by pairs of identical pions. 
Misidentified particles 
lead to counting of unlike pairs which do not give rise to 
Bose-Einstein correlation. It has been shown in~\cite{NA35a,NA35b} 
(NA35 exp.) and in~\cite{NA49} (NA49 exp.) 
that the main effect of particle misidentification is to reduce the 
value of the chaoticity parameter $\lambda$, which, however, is not 
used for the reconstruction of the size and dynamical state 
of the source; the relevant source parameters are effected 
by a few percents only. 
Moreover in this paper we present a compilation of recent results 
from other SPS experiments, which use either 
unidentified  $h^-$-$h^-$ pairs (NA49~\cite{NA49}, WA98~\cite{comp:WA98})
or $\pi^-$-$\pi^-$ pairs (NA44~\cite{comp:NA44}); their agreement 
supports the assumption that HBT correlation can be performed using  
$h^-$-$h^-$ pairs.
\par
The paper is structured as follows: the WA97 apparatus is briefly 
described in section~\ref{sec_0}. The experimental derivation of the 
two-particle correlation function is discussed in section~\ref{sec_1}; 
section~\ref{sec_CC} deals with the Coulomb corrections. 
In section~\ref{sec_2} the measured radii are presented 
as a function of the pair momentum; from them 
the source freeze-out size and temperature, the 
average freeze-out time and duration, and the expansion velocity 
are derived as functions of the collision centrality. 
From these observables a dynamical picture of the collision 
is depicted in section~\ref{sec_disc}. 
Finally conclusions are drawn in section~\ref{sec_last}.
\section{The WA97 experiment}
\label{sec_0}
\par
\begin{figure}[hbt]
  \centering
  \resizebox{0.38\textwidth}{!}{%
 \includegraphics{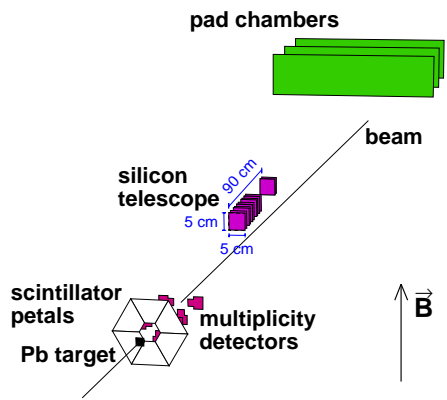}}
\caption{\rm The WA97 set-up.}
\label{fig:setup}
\end{figure}

The WA97 set-up, shown schematically in fig.~\ref{fig:setup}, is described
in detail in ref.~\cite{Alex95}. The target and the silicon telescope were
placed inside the homogeneous 1.8 T magnetic field of the CERN Omega
magnet. 
The 158~A~GeV/$c$ lead beam from the CERN SPS was incident on a lead target
with a thickness corresponding to 1\% of an interaction length.
Scintillator petal detectors behind the target provided an interaction
trigger selecting  approximately the most central $40\%$ 
of the Pb-Pb collisions.
Two planes of microstrip multiplicity detectors covering, for all
$p_T$\ values, the pseudorapidity
region $2 \lesssim \eta \lesssim 3$  and
$3 \lesssim \eta \lesssim 4$ respectively,
provided information for more detailed off-line study
of the centrality dependence of single and double particle spectra. 
In the present analysis the WA97 usual four centrality bins,  
described in  ref.~\cite{FL,WA97Centr} for the study of strangeness 
production, have been adopted.
\par
The core of the WA97 set-up is a silicon telescope consisting of 7 planes
of pixel detectors~\cite{25} which have a pixel size of 
$ 75 \, \times \, 500 \, {\rm \mu m^2}$\ and of 10 planes of 
$50 \, {\rm \mu m}$\ pitch silicon microstrips.
The telescope has $5 \, \times \,  5 \, {\rm cm^2}$\ cross section and  
contains $\approx \; 0.5 \,\times \, 10^6$\ detecting elements. 
This precise tracking device 
was placed $60 \, {\rm cm}$\ behind the target slightly above the 
beam line and inclined by $48 \, {\rm mrad}$, pointing to the target.
\par
The track reconstruction is done in the compact part of the telescope, 
where 11 planes of silicon detectors are closely packed over a 
distance of $ 30\,  {\rm cm}$. The momentum resolution for fast tracks is 
improved using lever arm detectors, which consists of additional pixel 
and microstrip planes and of three MWPC's with a cathode pad readout. \\
Due to the precise tracking and to its compactness 
the WA97 experiment is well suited for the measurements 
of the two-particle 
correlation functions. 
\section{Data analysis}
\label{sec_1}
\par
The quantum statistical correlation between two identical 
particles coming from a chaotic source yields information on the 
dimensions and the dynamics of the source. Experimentally the  
correlation is studied through a ``correlation function'', defined as: 
\begin{equation}
C_2(q)=\mathcal{N}
 \frac{S(q)}{B(q)}
\label{C2}
\end{equation}
where $S(q)$\ is the measured two-particle (Signal) distribution as a 
function of the relative four momentum $q=p_1-p_2$\ and $B(q)$\ is the 
reference (Background) distribution built by pairing two particles taken 
from different events. For each actual pair in one event (in the Signal) 
about 50 background pairs have been formed from events of similar 
multiplicity, so that the statistical error on $C_2$\ is dominated by 
the signal. 
The normalization factor $\mathcal{N}$\ is determined by imposing 
that the integral of $C_2(q)$\ over a wide interval outside the region of 
sensitive correlation (which is at low $q$) is equal to one. 
\par
From the correlation function $C_2$\ the description of the source 
has been extracted using two different parametrizations. 
{\begin{itemize}
\item[(1)] The ``standard'' Cartesian (or Pratt-Bertsch) 
 parametri\-zation~\cite{BP,ref17} is expressed in the 
{\em out-side-longitudinal} (OSL) system; this has the {\em longitudinal} 
direction along the beam axis, while in the transverse plane the {\em out} 
direction is chosen parallel to the transverse component of the pair 
momentum $\vec{K}_t$, the remaining Cartesian component denoting the 
{\em side} direction.
This parametrization is given by the following formula: 
\begin{equation}
C_2(q)=  1+\lambda \exp\left[-R^2_oq^2_o-R^2_sq^2_s 
         -R^2_lq^2_l-2|R_{ol}|R_{ol}q_lq_o \right]
\label{BP}
\end{equation}
where $q_o$,  $q_s$,  $q_l$\ are the spatial components of $q$\ in 
the OSL system and 
$R_o$, $R_s$, $R_l$, $R_{ol}$, 
which are fit parameters, 
provide information on the extension and dynamics of the source~\cite{ref17}. 
$\lambda$\ is called the ``chaoticity parameter'' 
and is equal to unity for a completly chaotic source and smaller than 
unity for partially chaotic source. 
It is however affected in a non trivial way by resonance decays, 
particle misidentification and other effects~\cite{Boal}, 
which are expected to depend on the pair momentum. 
\item[(2)]
The Yano-Koonin-Podgoretskii (YKP) parametrization \cite{YKP} 
uses the components $q_{\bot}=\sqrt{q_o^2+q_s^2}$, $q_0=E_1-E_2$, 
$q_{\parallel}=q_l$\ and starts from the ansatz: 
\begin{equation}
C_2(q) = 1+\lambda \exp \left[ 
 -R^2_{\perp}q^2_{\perp}-  
   \gamma^2_{yk}(q_{\parallel}-
   v_{yk}q_{0})^2 R^2_{\parallel} 
   -\gamma^2_{yk}(q_{0}-v_{yk}q_{\parallel})^2R^2_0 \right] 
\label{YKP}
\end{equation}
where $R_{\perp}$, $R_{\parallel}$, $R_{0}$\ are called the ``YKP radii'', 
although $R_0$\  has temporal dimension (${\rm fm}/c$);
$v_{yk}$, called  ``Yano-Koonin velocity'', 
is measured in units of {\em c} and 
$\gamma_{yk}=(1-v_{yk}^2)^{-1/2}$.
\end{itemize}
\par
The transverse orientation of the reference system is 
different in the two parametrizations. 
Concerning the longitudinal direction, the LCMS 
({\em longitudinal co-moving system}) frame, in which the average pair 
momentum has only a transverse component ($K_l=0$), has been chosen 
for both of them. 
This system has the advantage of a more straightforward interpretation 
of the source parameters. 
Hence, a reference system is used where  
the relative momentum $q$ is measured on a pair-by-pair basis. 
The same is true 
in the transverse direction for the Cartesian parametrization. 
\par
As claimed in ref.~\cite{E877}, fitting by the Least Square method and using 
the square root of the number of counts as an estimator of the error 
introduces a systematic bias. A Maximum Likelihood method has been applied 
by minimizing the negative logarithmic likelihood function: 
\begin{equation}
-2\ln{\mathcal{L}(\vec{R})}=2\sum_i \left[
C_iB_i-S_i \ln(C_iB_i)+\ln(S_i!) \right]
\nonumber
\end{equation}
where $C_i$ is the correlation hypothesis for a given set of fit parameters 
$\vec{R}$\ (e.g. $\vec{R}=(\lambda,R_{\perp},R_{\parallel},R_0,v_{yk})$).
\par 
The main quality test of the full analysis procedure 
comes from 
the consistency between parameters 
obtained from Cartesian and YKP forms of the correlation function. 
This topic is discussed in section~\ref{sec_2}. 
Two secondary checks on the fit quality have been performed. 
Firstly, the $\chi^2/{\rm NDF}$--values are calculated for the 
parameters corresponding to the best likelihood fits and are 
found to be distributed around unity. Secondly, we have fitted the projections 
of the 3-dimensional correlation functions onto one momentum 
difference component, with narrow cuts on the other components, 
using a least-squares method to a Gaussian function. This yields  
results consistent with the results of the 
3-dimensional fits. A typical sample of these projections is 
shown in fig.~\ref{fig:projec} 
where the correlation function has been 
integrated  along the other $q_j$\ components 
in the intervals $| q_j |  <  20\, {\rm MeV}$, 
except for the $q_0=E_1-E_2$\ projection where 
$|q_{j}| \, < \, 50\, {\rm MeV}/c\;  (j=\perp,\parallel)$.
\begin{figure}[t]
  \centering
\resizebox{0.48\textwidth}{!}{%
\includegraphics{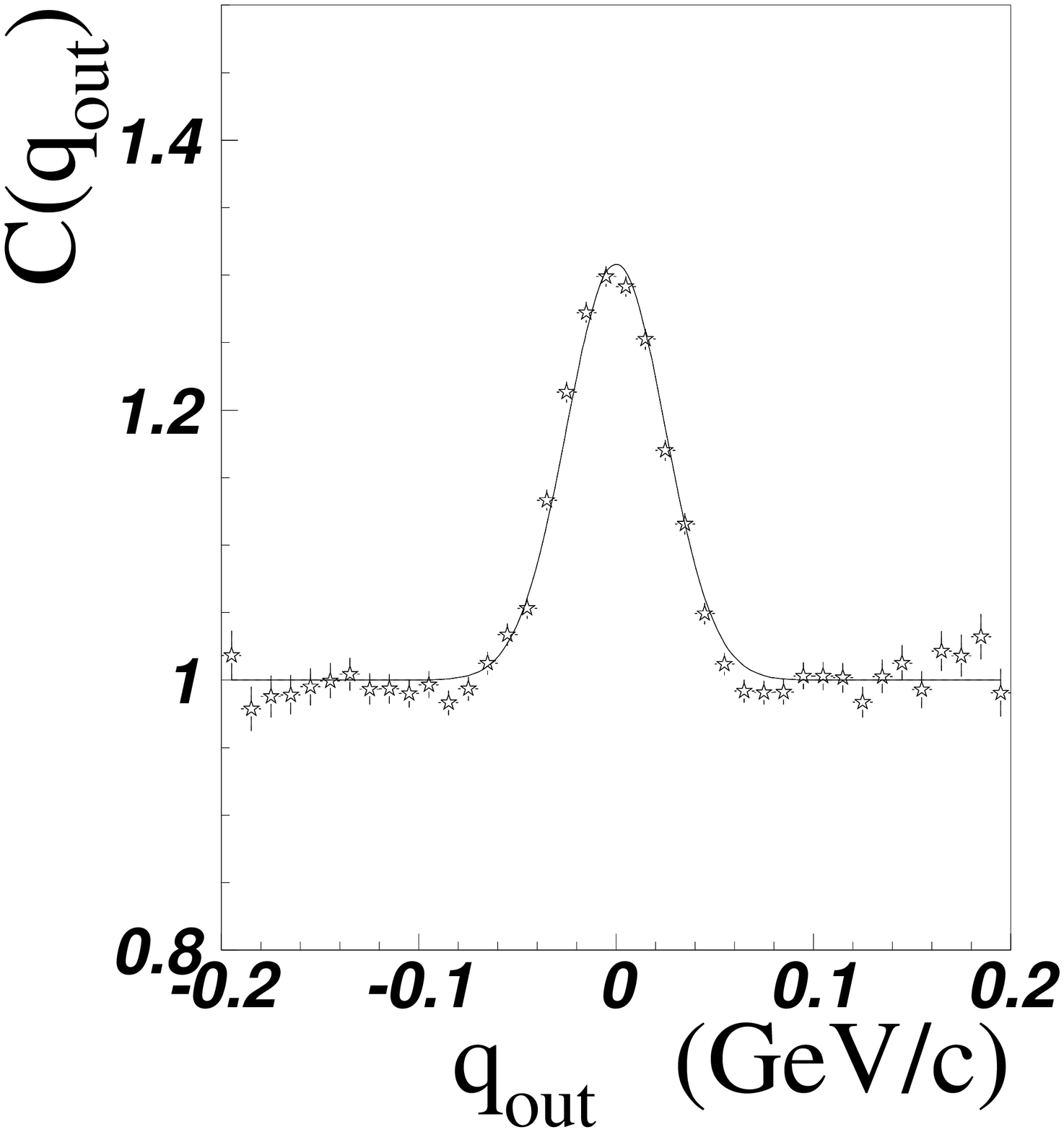}
\includegraphics{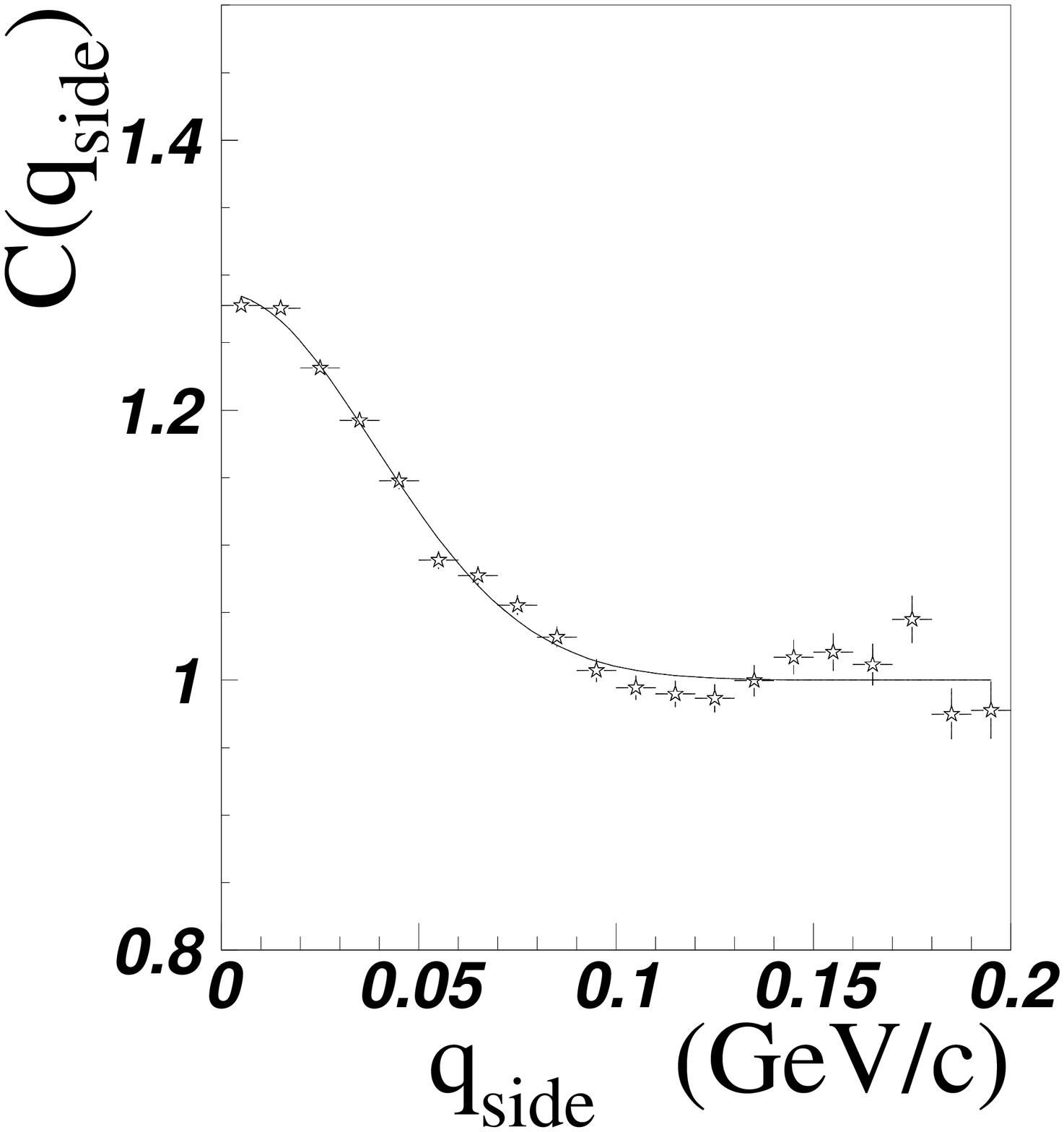}
\includegraphics{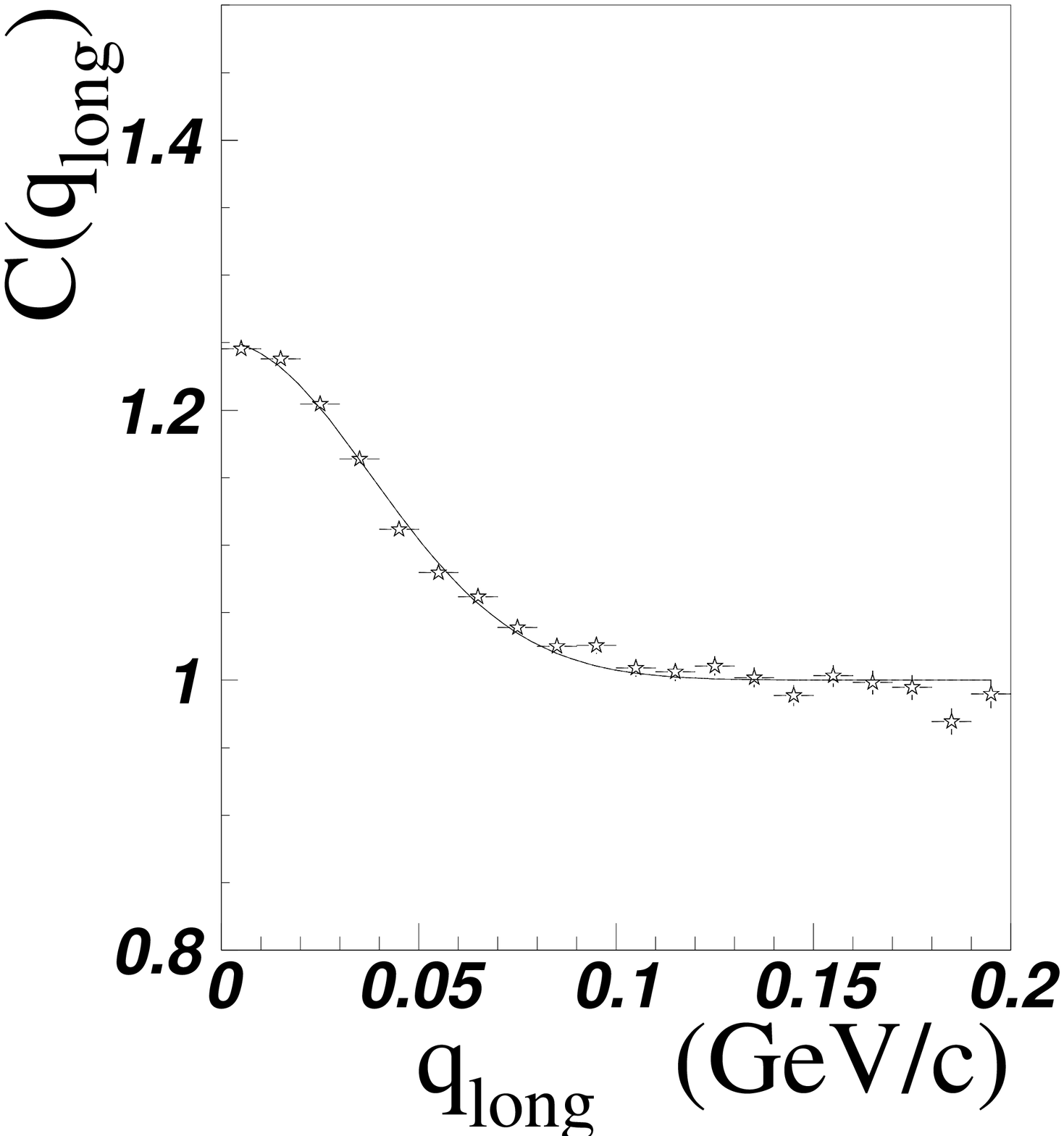}}\\
\resizebox{0.48\textwidth}{!}{%
\includegraphics{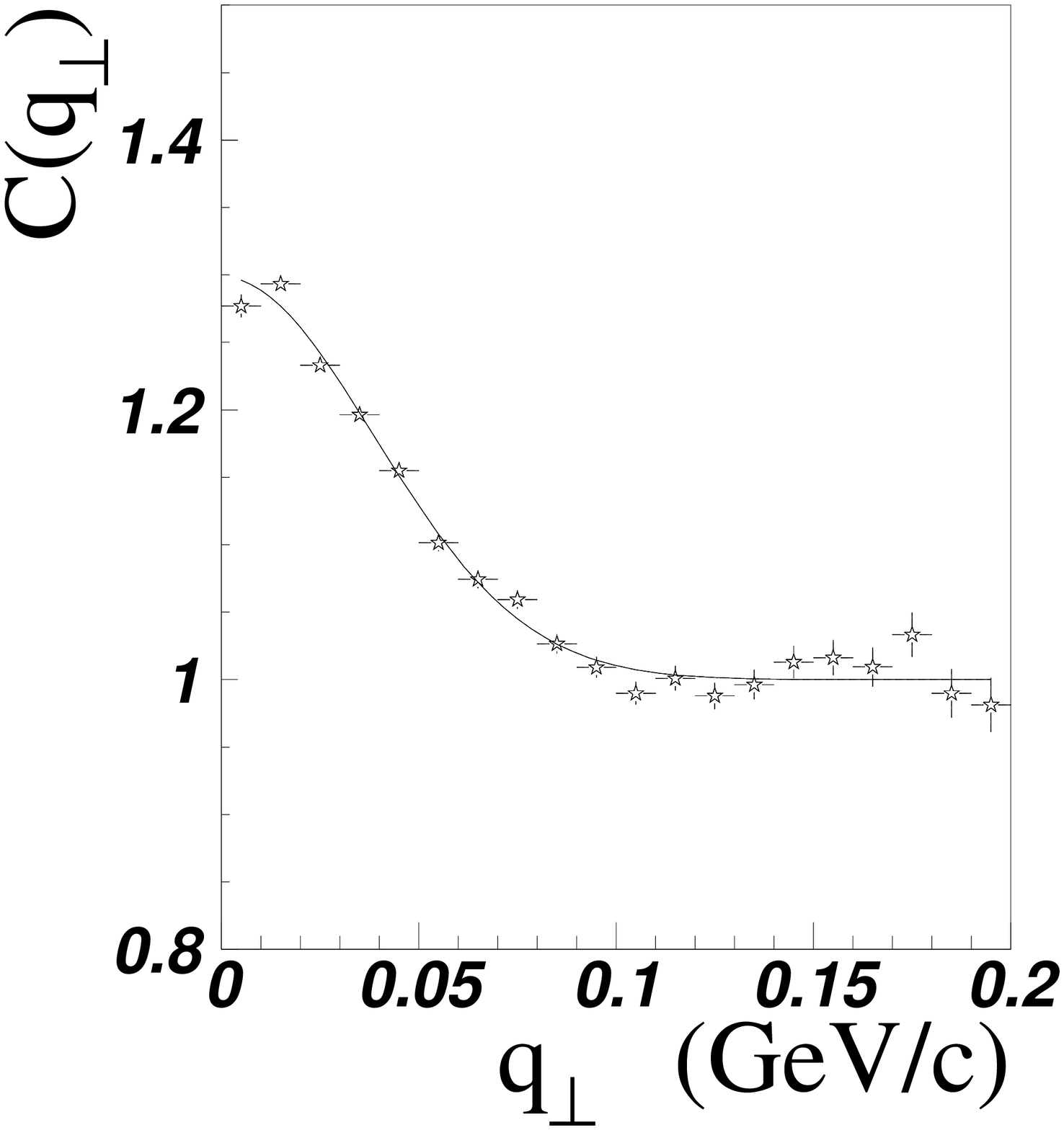}
\includegraphics{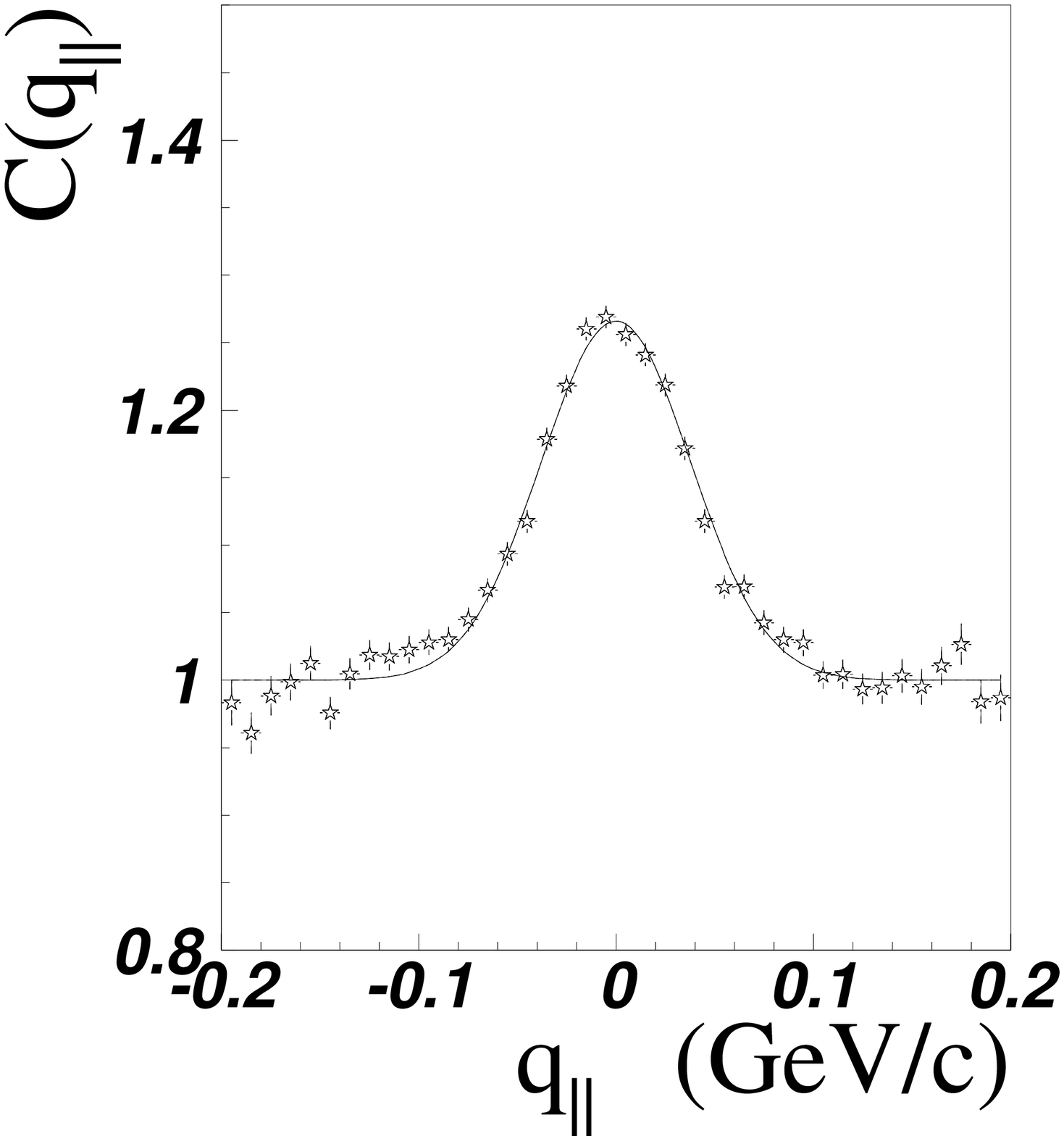}
\includegraphics{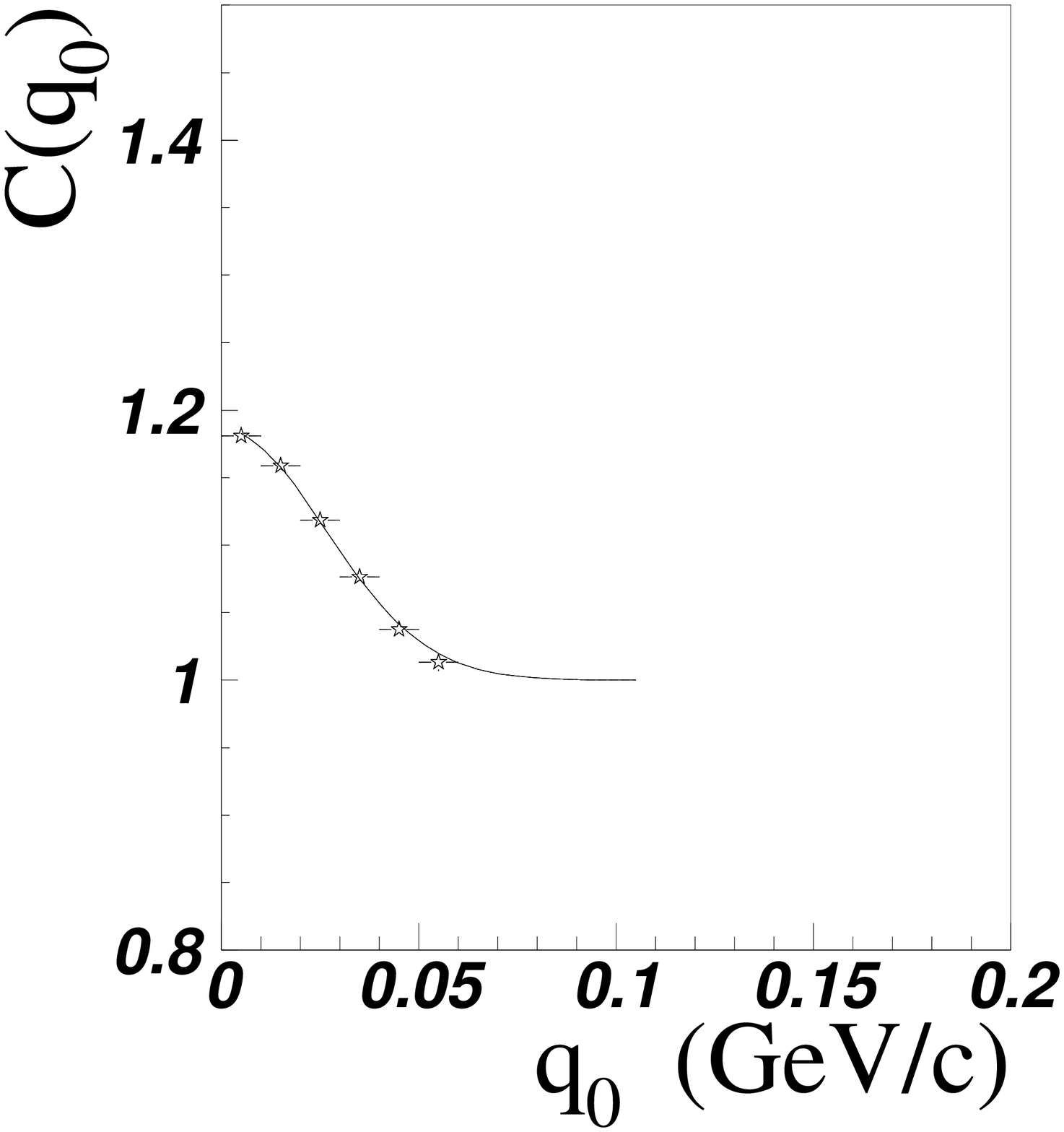}}
\caption{{\rm Projections of the $h^-$-$h^-$\ correlation function measured
using Cartesian (top) and YKP (bottom) parametrization for pair 
rapidities $ 2.55  <  Y_{\pi\pi}  <  3.65 $\ and pair trasverse 
momenta $ 0.23  <  K_t  <  0.39 \, {\rm GeV}/c$. 
Solid curves are the result of a least-square fit to a 
Gaussian function (see the text).}}
\label{fig:projec}
\end{figure}
\par
The determination of the size and dynamical state of the source at 
freeze-out requires information about the dependence of the YKP radii 
on the mean momentum $\vec{K}=\frac{1}{2}(\vec{p}_1+\vec{p}_2)$ of the pair. 
This can be parametrized by its transverse component 
$K_{t}=\frac{1}{2}\sqrt{(p_{y1}+p_{y2})^2+(p_{z1}+p_{z2})^2}$\ 
and the pair rapidity 
$Y_{\pi\pi}=\frac{1}{2}\log\frac{E_1+E_2+p_{x1}+p_{x2}}
{E_1+E_2-p_{x1}-p_{x2}}$. 
With this in mind, our (two-particle) acceptance window, which 
is shown in fig.~\ref{fig:accept}, 
has been binned in 20 rectangles in each of which  
the correlation function $C_2(q)$\ has been measured independently.
\begin{figure}[t]
  \centering
  \resizebox{0.42\textwidth}{!}{%
  \includegraphics{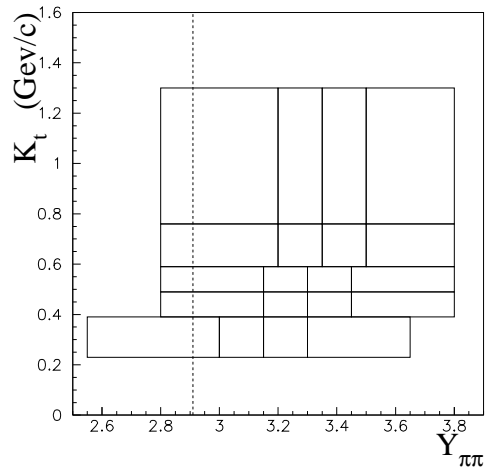}}
\caption{{\rm Two-particle acceptance window (assumed to be pions)
 used in the present analysis. 
 The correlation function has been measured in each of the 
 twenty rectangles patching the window together. 
 The $K_t$\ and $Y_{\pi\pi}$\ 
 average values, measured over the rectangles   
 are reported in table~1. The dashed line is drawn 
 at $Y_{\pi\pi}=y_{cm}$.}} 
\label{fig:accept}
\end{figure}
Table~1 displays the average values of 
the $K_t$\ and $Y_{\pi\pi}$\ distributions in each of the 20 bins used 
for the analysis. \vspace{0.6cm} \\
{\footnotesize {\bf Table 1. } Averages of the 
 $K_t$\ (uppermost number in the cells, in ${\rm GeV}/c$\ units) and 
 $Y_{\pi\pi}$\ (bottom numbers) distributions  in the 20 
 rectangles of fig.~\ref{fig:accept}.}
\begin{center}
\begin{tabular}{||l|c|c|c|c||} \hline \hline
\multicolumn{1}{|c|}{ } & \multicolumn{4}{c|}
{\bf $Y_{\pi\pi} \, \longrightarrow$} \\ \hline
& $\begin{array}{cc} 0.910 \\ 3.11 \end{array} $ &
  $\begin{array}{cc} 0.915 \\ 3.28 \end{array} $ &
  $\begin{array}{cc} 0.915 \\ 3.42 \end{array} $ &
  $\begin{array}{cc} 0.925 \\ 3.60 \end{array} $  \\ \cline{2-5}
& $\begin{array}{cc} 0.660 \\ 3.10 \end{array} $ &
  $\begin{array}{cc} 0.665 \\ 3.28 \end{array} $ &
  $\begin{array}{cc} 0.665 \\ 3.42 \end{array} $ &
  $\begin{array}{cc} 0.670 \\ 3.60 \end{array} $  \\ \cline{2-5} 
{\bf $\uparrow $} &
  $\begin{array}{cc} 0.535 \\ 3.05 \end{array} $ &
  $\begin{array}{cc} 0.535 \\ 3.23 \end{array} $ &
  $\begin{array}{cc} 0.540 \\ 3.37 \end{array} $ &
  $\begin{array}{cc} 0.540 \\ 3.56 \end{array} $  \\ \cline{2-5}  
{\bf $K_t$}  &
  $\begin{array}{cc} 0.435 \\ 3.03 \end{array} $ &
  $\begin{array}{cc} 0.440 \\ 3.23 \end{array} $ &
  $\begin{array}{cc} 0.440 \\ 3.37 \end{array} $ &
  $\begin{array}{cc} 0.445 \\ 3.55 \end{array} $  \\ \cline{2-5} 
& $\begin{array}{cc} 0.315 \\ 2.86 \end{array} $ &
  $\begin{array}{cc} 0.330 \\ 3.08 \end{array} $ &
  $\begin{array}{cc} 0.335 \\ 3.22 \end{array} $ &
  $\begin{array}{cc} 0.345 \\ 3.42 \end{array} $  \\ \hline \hline
\end{tabular}
\label{tab:accept}
\end{center}
\vspace{0.6cm}

Additionally, in order to investigate the dynamics of the collisions 
as a function of centrality, the whole analysis has been repeated 
for each of the four centrality classes selected by WA97 to 
study the strange particle  
production~\cite{FL,WA97Centr}.
\par 
The quality of a correlation measurement is determined by the 
resolution of the two-particle momentum difference 
(``relative momentum'') $q$. 
The relative momentum resolution was studied using a 
Monte Carlo chain based on the 
simulation code GEANT~\cite{ref:GEANT}. As reported in table~2, 
we have found that all relative momentum projections used in this analysis 
($q_{o},\,q_{s},\,q_{l};\, q_{\perp},\,q_{\parallel},\,q_{0}$) 
are measured with an error not bigger than $10 \, {\rm MeV}/c$\ 
over our two-particle acceptance window. \vspace{0.6cm} \\
{\footnotesize {\bf Table 2.} 
 The resolution of relative momenta $q_{o},q_{s},q_{l}$\ and 
 $q_{\bot},q_{\parallel},q_{0}$ in the LCMS frame.}
\begin{center}
\begin{tabular}{||c|c||}  \hline
 Projection & Resolution (MeV/c)\\
\hline\hline
$ q_{o} $        &  $ 7.5 $  \\ \hline
$ q_{s}$         &  $ 10  $  \\ \hline
$ q_{l}$         &  $ 7.5 $  \\ \hline\hline
$ q_{\bot}  $    &  $ 9.0 $  \\ \hline
$ q_{\parallel}$ &  $ 7.5 $  \\ \hline
$ q_{0}   $       &  $ 7   $ \\ \hline\hline
\end{tabular}
\end{center}
\vspace{0.6cm}
The bin size used to build the correlation functions 
has been accordingly taken to be $10 \, {\rm MeV}/c$\ in all 
relative momentum components. 
It has been shown in~\cite{ref:Appel} that due to the finite bin size 
the values of the extracted HBT radii are systematically 
underestimated, the greater the radius the higher the error. 
Since the typical half-width of the correlation function 
in our acceptance window is $\sim 50 \, {\rm MeV}/c$\  
(corresponding to a typical radius of about $4\, {\rm fm}$), 
our binning yields an underestimation smaller than 10\%. 
\par
The detector's ability to distinguish a pair of close tracks from 
a single track decreases with decreasing track separation, depending 
on the Si-pixel size. 
The correction for this inefficiency 
was calculated via a Monte Carlo simulation 
by reconstructing 
a sample of (generated) two-tracks events embedded in real ones. 
The maximum inefficiency was found to be $ \approx  4\%$\ for 
pairs populating the bin of the correlation functions 
centred at $q=0$ (relative to high $q$\ pairs).
\par
The results presented here are based on 17 million events collected 
in the 1995 run. Only tracks pointing to the main interaction vertex, 
which is reconstructed on a event by event basis, are selected. 
After event reconstruction we end up with 13 million $h^-$\ pairs 
in the correlation sensitive region $ |\vec{q}| \, < \, 100 \, {\rm MeV}/c$\ 
(LCMS system). 
\section{Coulomb correction}
\label{sec_CC}
\par
The observed two-particle correlation is the result of two 
different contributions, the Bose-Einstein effect and 
the Coulomb interactions. 
The Coulomb interaction between 
the particles of a pair 
accelerates them relative to each other, thus depleting the 
two-particle correlation function at small relative momenta.
\par
In order to get the Bose-Einstein correlation alone, 
the correlation function $C_2$\ needs to be 
corrected for the Coulomb interaction among the particles, so that:
\begin{equation}
C_2(p_1,p_2)=C_2^{raw}(p_1,p_2) \times K_{Coul}(p_1,p_2)
\end{equation} 
where $C_2^{raw}$\ is the B-E correlation of the experimental data and 
$K_{Coul}$\ is the correction for the Coulomb interaction.
In the past, the correction was done by describing the emission of an 
isolated pairs of charged particles from a point-like neutral region 
(Gamow factor~\cite{Gamov,Gyulassy}). This correction  
has been shown to be inadequate in heavy ion collision 
at the SPS energy~\cite{Gamovbad1,Gamovbad2}, 
essentially due to the extent of the source which is far 
from a point-like approximation. 
The Gamow factor has been considered in this analysis just as a reference 
for the two correction methods implemented. Those are: 
\begin{itemize}
\item[i)]
The method based on the measure of oppositely charged particle correlation, 
following the approach of Baym and Braun-Munzinger~\cite{Baym}.\\ 
Oppositely charged particles do not present correlation due to 
quantum statistical effects; they can thus provide a direct measure of 
the additional correlation introduced by the Coulomb interaction, 
accounting for the reverse in the charge sign. 
The method suggested in~\cite{Baym} is a classical model 
which neglects Coulomb interaction inside a pair for separations 
less then a given $r_0$\ and includes it for larger separations. 
A charge screening effect, expected in the high multiplicity environment 
created just after an ultra-relativistic heavy ion collision, 
is thus taken into account. 
The only free parameter of the model, $r_0$, can be obtained by 
a fit to the measured data points of the correlator for oppositely 
charged particles. The obtained value, $r_0=6.1 \pm 0.5 \, {\rm fm}$, 
is consistent with the NA49 measurement~\cite{NA49}. 
The resulting correction function 
is shown in fig.~\ref{fig:compFC}  as a function of 
$q_{inv}=\sqrt{\vec{q}^2-q_0^2}$. \\
\item[ii)]
The full Coulomb wave-function integration~\cite{Pratt1,Pratt2}.
This method has the advantage of being 
adaptable to any emission mechanism and can be implemented in bin-by-bin 
correction, independently of 
which particular projections 
of $q=p_1-p_2$\  are in use. Coulomb interactions with the residual 
nuclear system are neglected.
As a validation test of the procedure, we first assumed a point-like 
source (i.e. very small volume, $<<\, 1\,{\rm fm}$) 
with sudden emission (i.e. the particles are emitted at the same time), 
comparing to the Gamow correction. The agreement is 
satisfactory 
as shown in fig.~\ref{fig:compFC}. 
Then two kinds of source have been implemented: 
\begin{itemize}
\item[(a)]
a Gaussian source with sudden emission, which has been traditionally 
used in connection with the wave-function integration. 
\item[(b)]
 the same source model adopted to investigate the features of the 
 expansion dynamics and geometry of the fireball.\\ 
Since the parameters of the model, namely 
$ {\rm T,\,\beta_{\perp},\,R_{G},\,\Delta\eta,\,\Delta\tau,\,\tau_0,} $\
are required to describe the source during the computation of the Coulomb 
correction, an iterative procedure has been followed:
\begin{itemize}
\item[-]
a first complete analysis of the 
correlation functions corrected with the Baym and Braun-Munzinger approach, 
leads to a first estimate of the model parameters; 
\item[-] these parameters 
are 
used to calculate a new Coulomb 
correction by means of the full wave-function integration method; 
\item[-] a second full analysis of the 
 correlation with the new Coulomb correction 
 (that will be presented in the next section) 
 leads 
 to a second, better evaluation of the model parameters. 
\end{itemize}
The procedure converged satisfactorily after the second iteration.
\end{itemize}
\end{itemize}
\par 
A final comparison of all the methods is shown in fig.~\ref{fig:compFC}. 
As expected, the Gamow factor significantly overestimates the Coulomb 
interaction in heavy ion collision; 
the corrections calculated with the static Gaussian source and those 
of the Baym model, that implicitly also assumes a static source, are found 
to be very similar. 
The expanding source has proven to yield the smallest corrections. 
We may conclude that treating a rapidly expanding source as if it were a 
{\em  static} system can lead to a slight overestimate of the 
Coulomb corrections.
\begin{figure}[hbt]
  \centering
  \resizebox{0.48\textwidth}{!}{%
\includegraphics{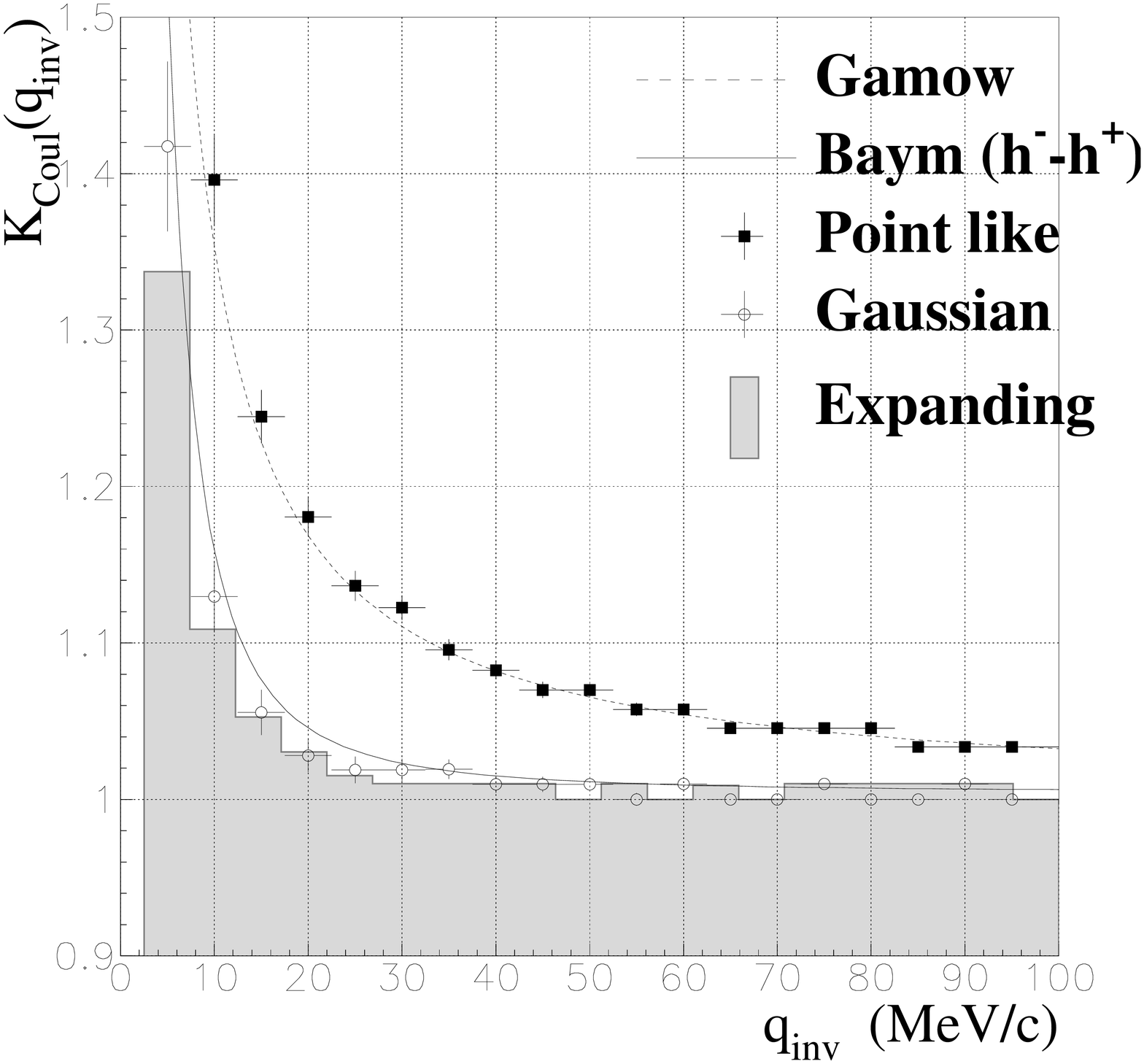}}
\caption{\rm Corrections to the correlation function
  for the Coulomb interaction according different models: 
 the Gamow factor (dashed line), 
 the Baym method based upon the correlation of 
  oppositely charged particles (solid line), 
 the full wave-function integration for a point-like source (black squares) 
  and for a (static) Gaussian source with rms radius of 
  $4\, {\rm fm}$\ (open circles), 
  the full wave-function integration for an expanding 
  thermalized source~\cite{RecProgr2} (shadow area).}
\label{fig:compFC}
\end{figure}
The correction has been applied on a bin-by-bin basis, 
acting on the $q_{inv}$\ of that bin, even if we are making a correlation 
function in some other set of variables. 
\section{Correlation study}
\label{sec_2}
\subsection{Cartesian and YKP parametrizations: consistency}
\par
In the context of the model by Chapman et al.~\cite{RecProgr} 
for a finite expanding 
thermalized source, the YKP parameters give more direct access to 
the features of the source. Therefore in this paper we shall 
follow this approach. 
However, the Cartesian radii have been systematically used to test the 
goodness of the fitting procedure of the YKP correlation function, 
by means of equations relating the two sets of parameters (for these 
equations see e.g.~\cite{RecProgr2,Consistency2}). 
Fig.~\ref{fig:comp} shows typical fit results for the parameters 
$(\lambda, R_{\perp}, R_{\parallel}, R_0)$\ versus $K_t$\   
as obtained directly from YKP parametrization (black triangles) and 
deduced from the corresponding Cartesian set of parameters (open circles); 
as a second representative sample, in fig.~\ref{fig:comp_v} are 
displayed the Cartesian radii versus $Y_{\pi\pi}$\ evaluated 
directly and from the corresponding YKP correlation functions. 
It is worth stressing that the analysis has been carried out in 
a completely independent way for the two parametrizations: this 
means, for instance, that the corrections 
for Coulomb interaction and two-track resolution have been 
calculated and applied independently. In short, 
Cartesian and YKP radii give two 
separate 
measurements of the same physical quantities; 
their full compatibility gives confidence on the 
absence of systematic errors in the analysis. 
\begin{figure}[hbt]
  \centering
\resizebox{0.44\textwidth}{!}{%
\includegraphics{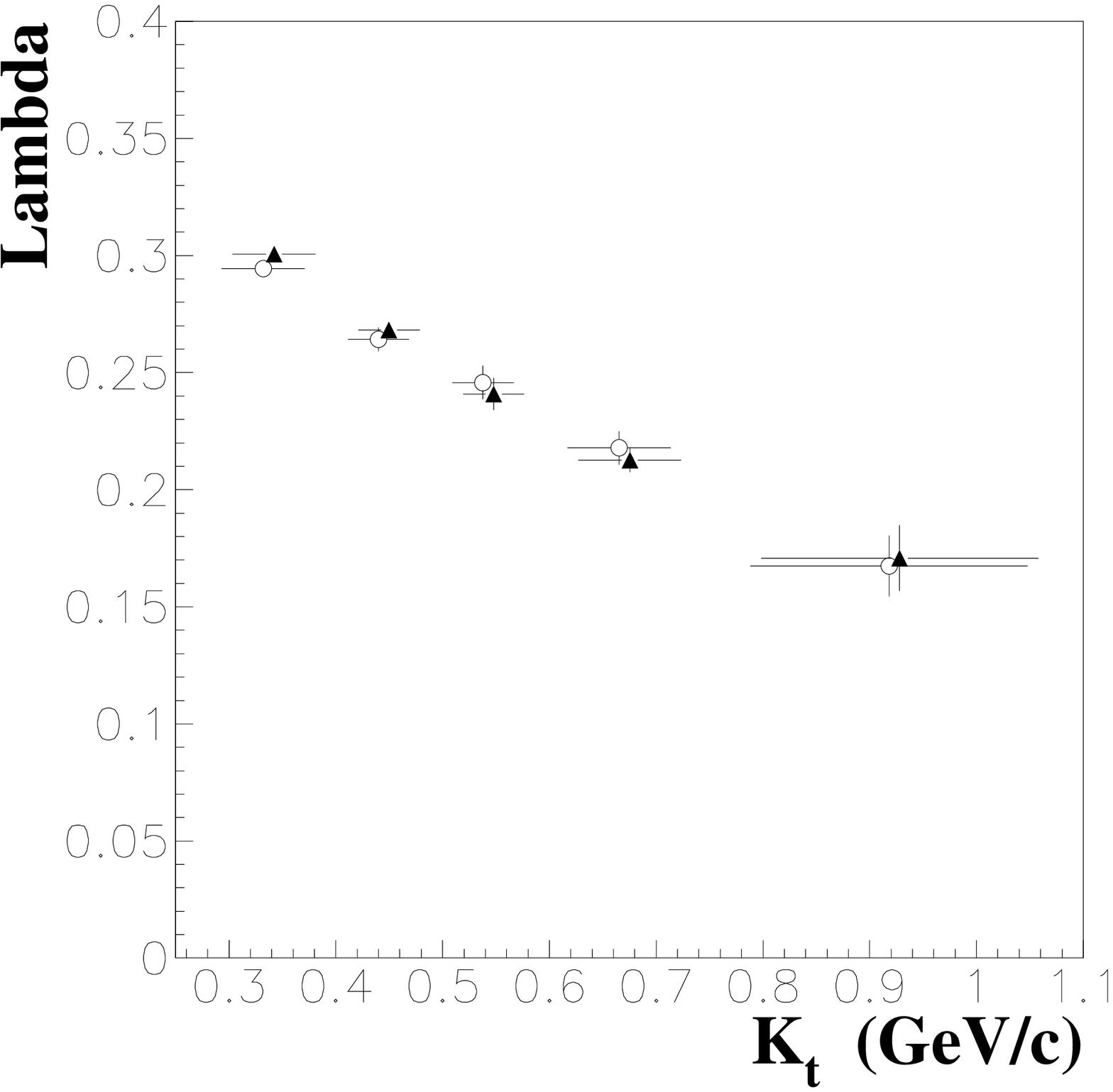}
\includegraphics{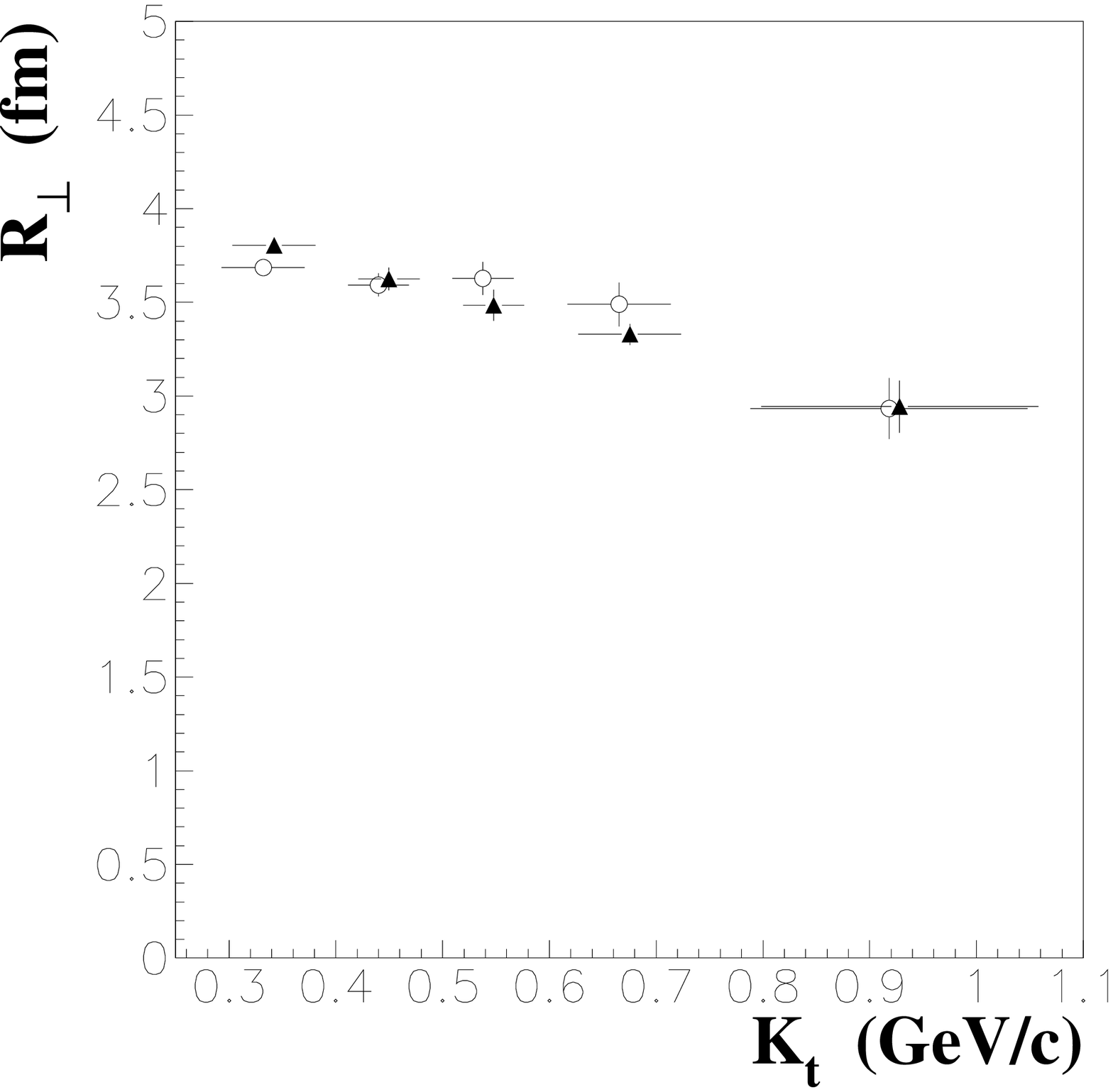}}\\
\resizebox{0.44\textwidth}{!}{%
\includegraphics{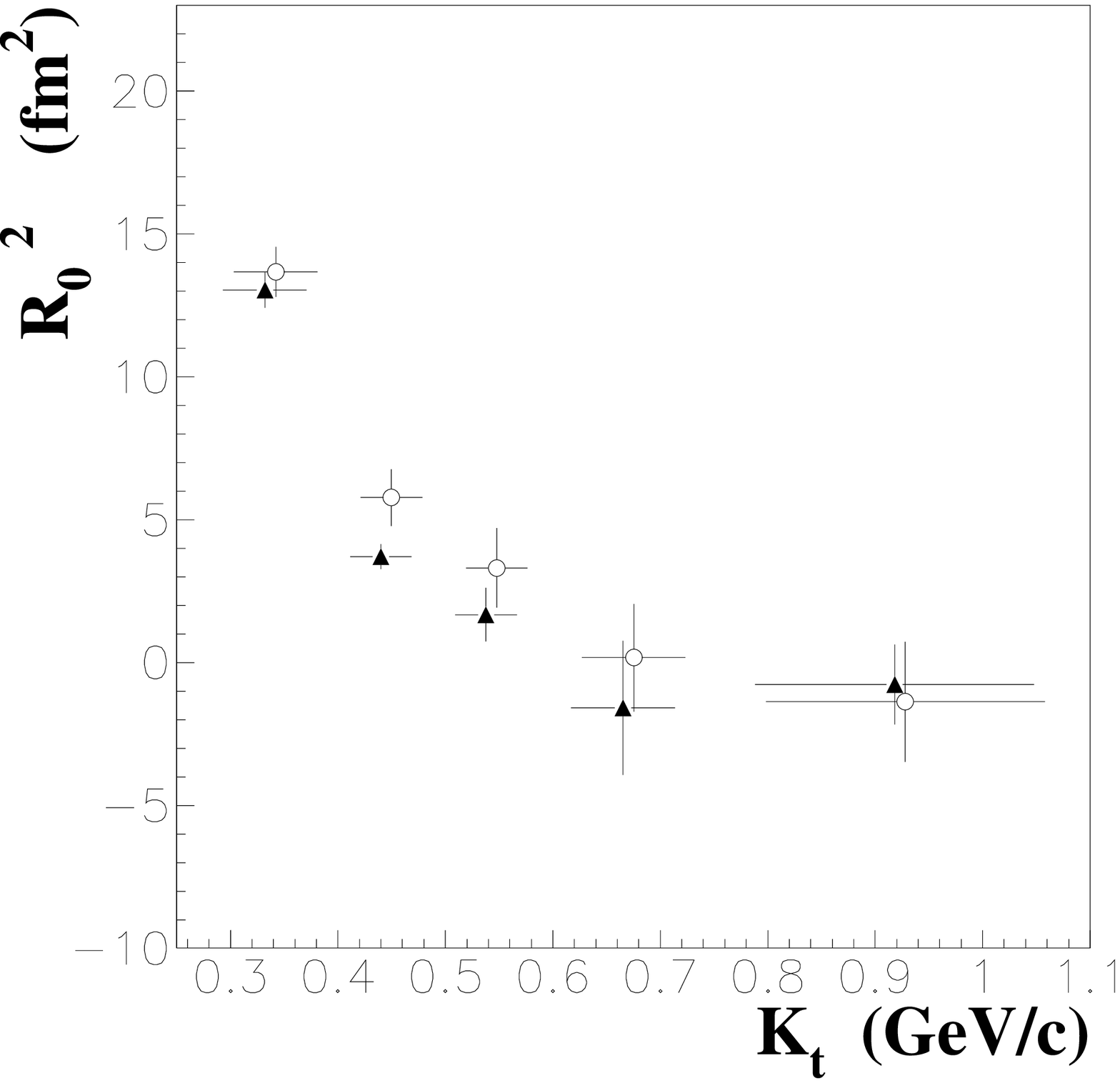}
\includegraphics{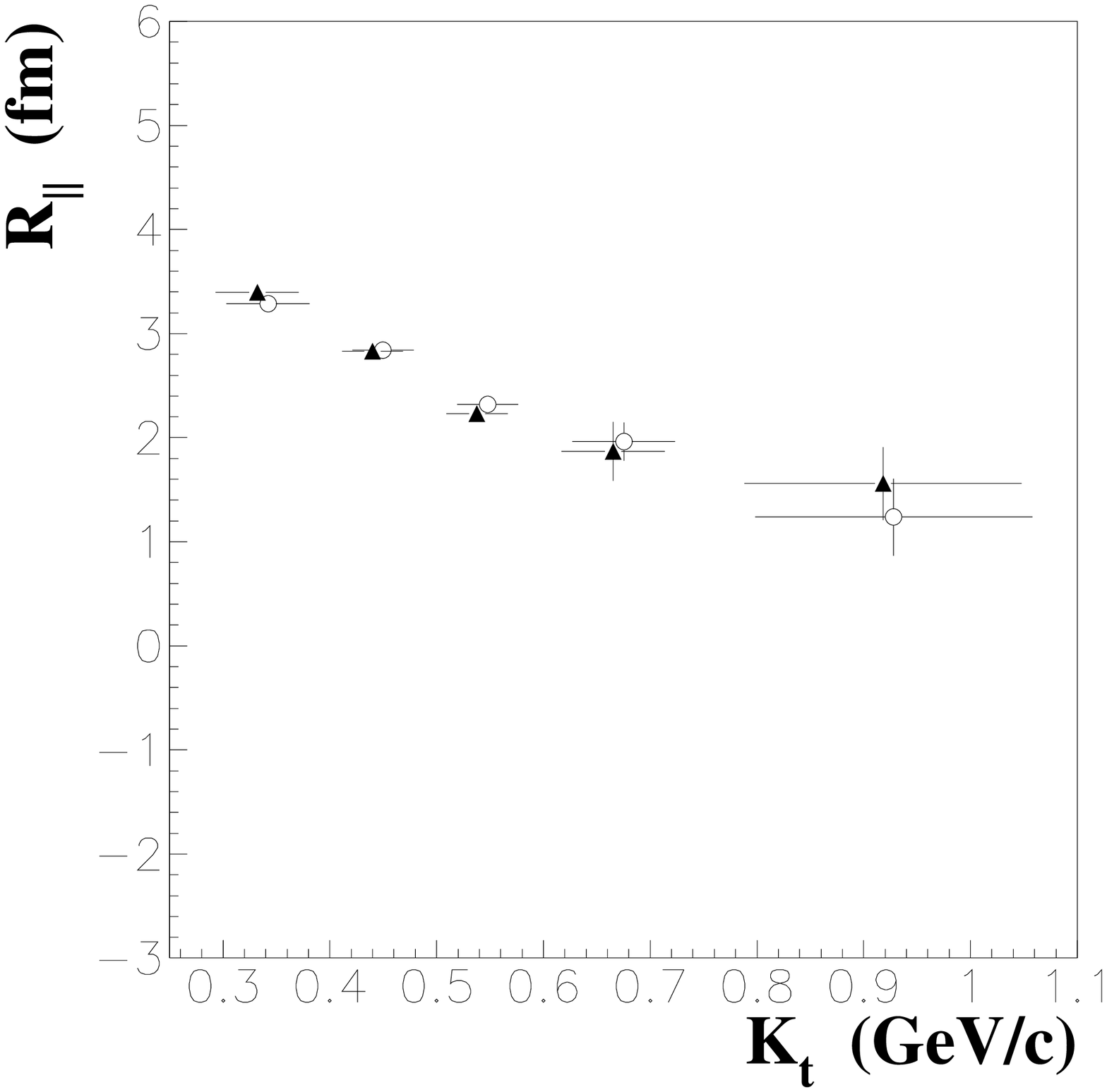}}
\caption{\rm The $\lambda$\ factor and the YKP radii as calculated 
directly (black triangles) and from the Cartesian parametrization 
 (open circles) as a function of $K_t$\ in the full 
 $Y_{\pi\pi}$\ and centrality interval. The remaining parameter 
 $v_{yk}$\ is displayed in fig.~\ref{fig:vyk2} 
 (plot labelled ``ALL'').}
\label{fig:comp}
\end{figure}
\begin{figure}[hbt]
  \centering
\resizebox{0.22\textwidth}{!}{%
\includegraphics{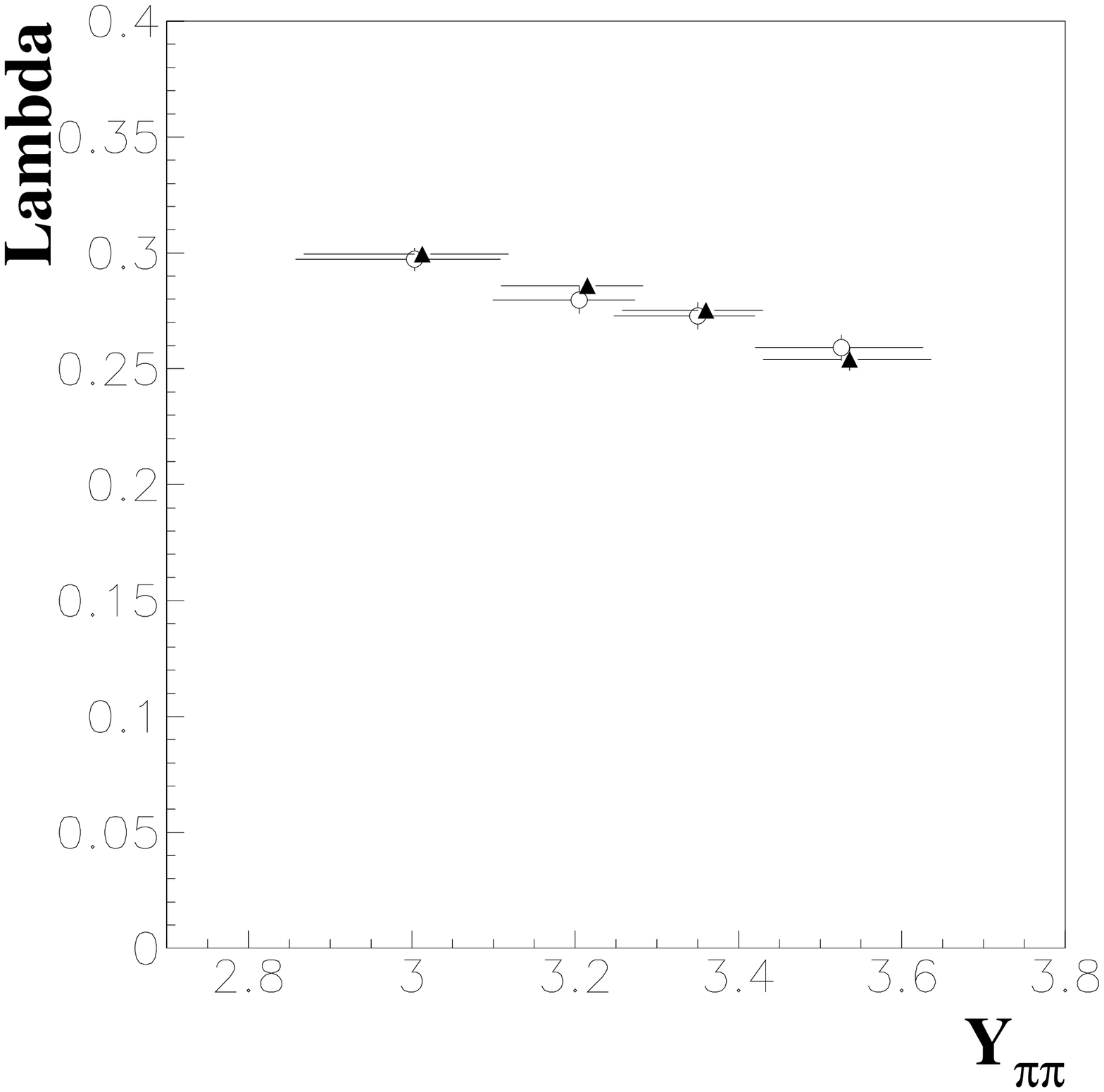}}\\
\resizebox{0.44\textwidth}{!}{%
\includegraphics{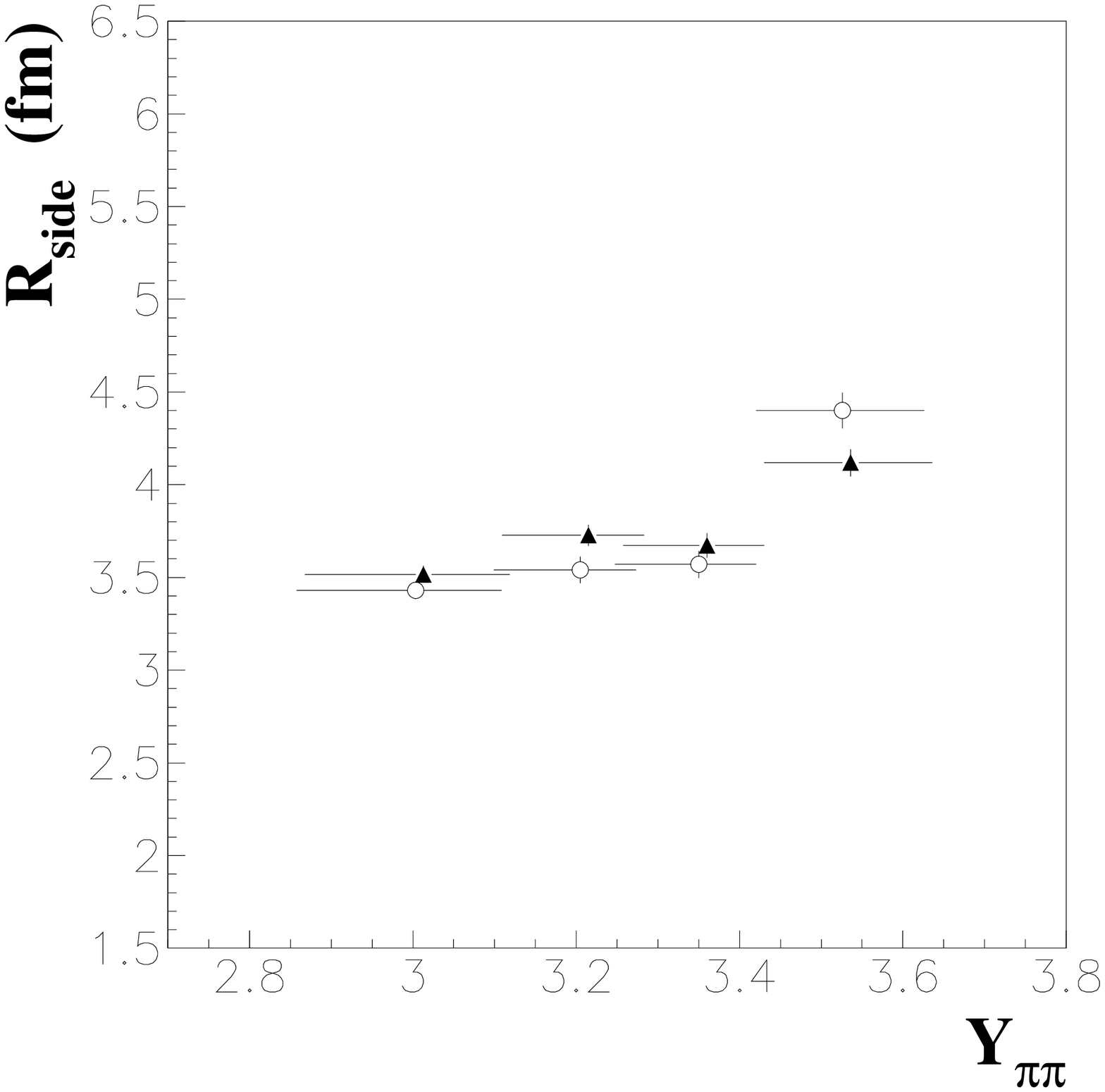}
\includegraphics{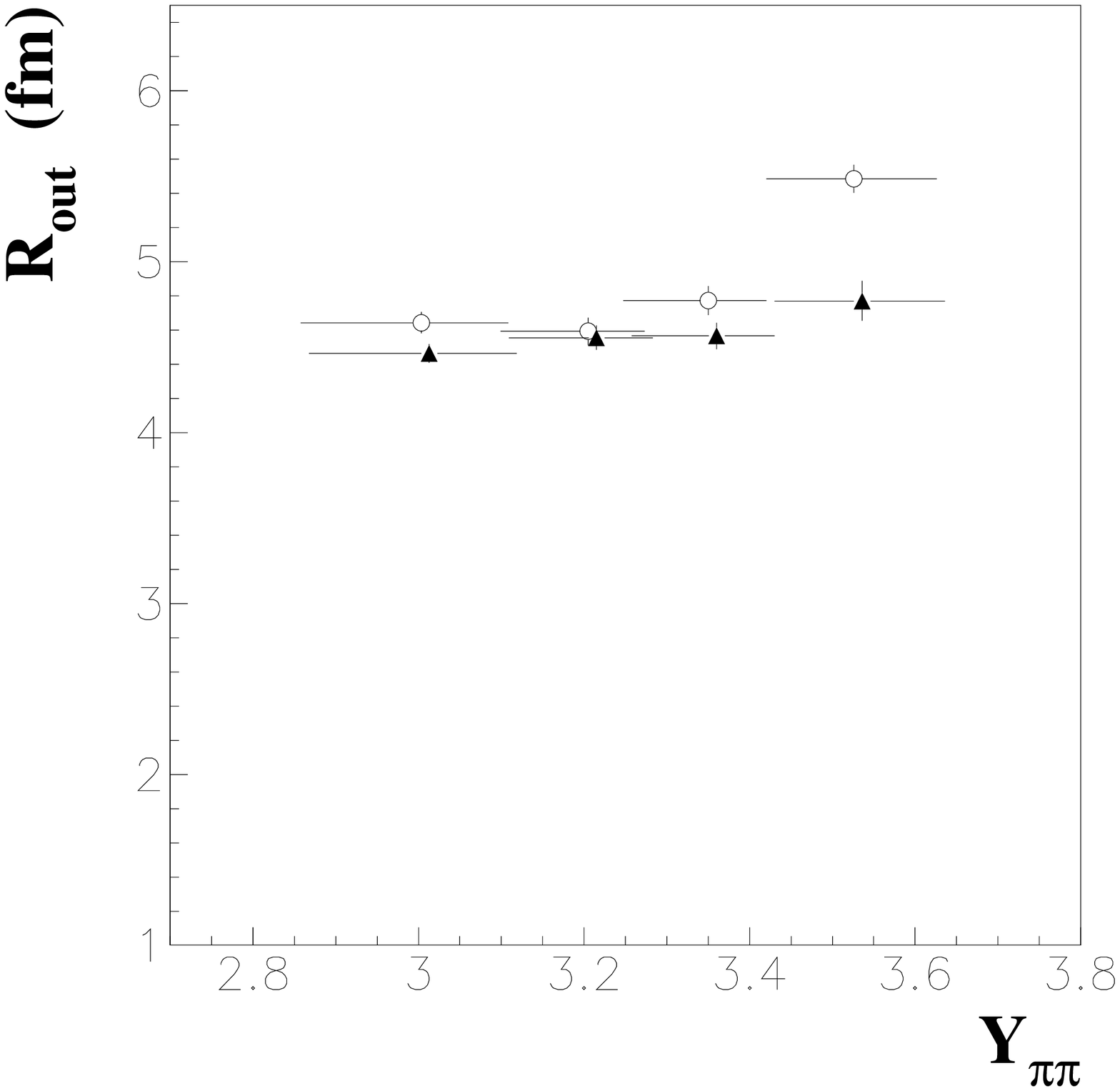}}\\
\resizebox{0.44\textwidth}{!}{%
\includegraphics{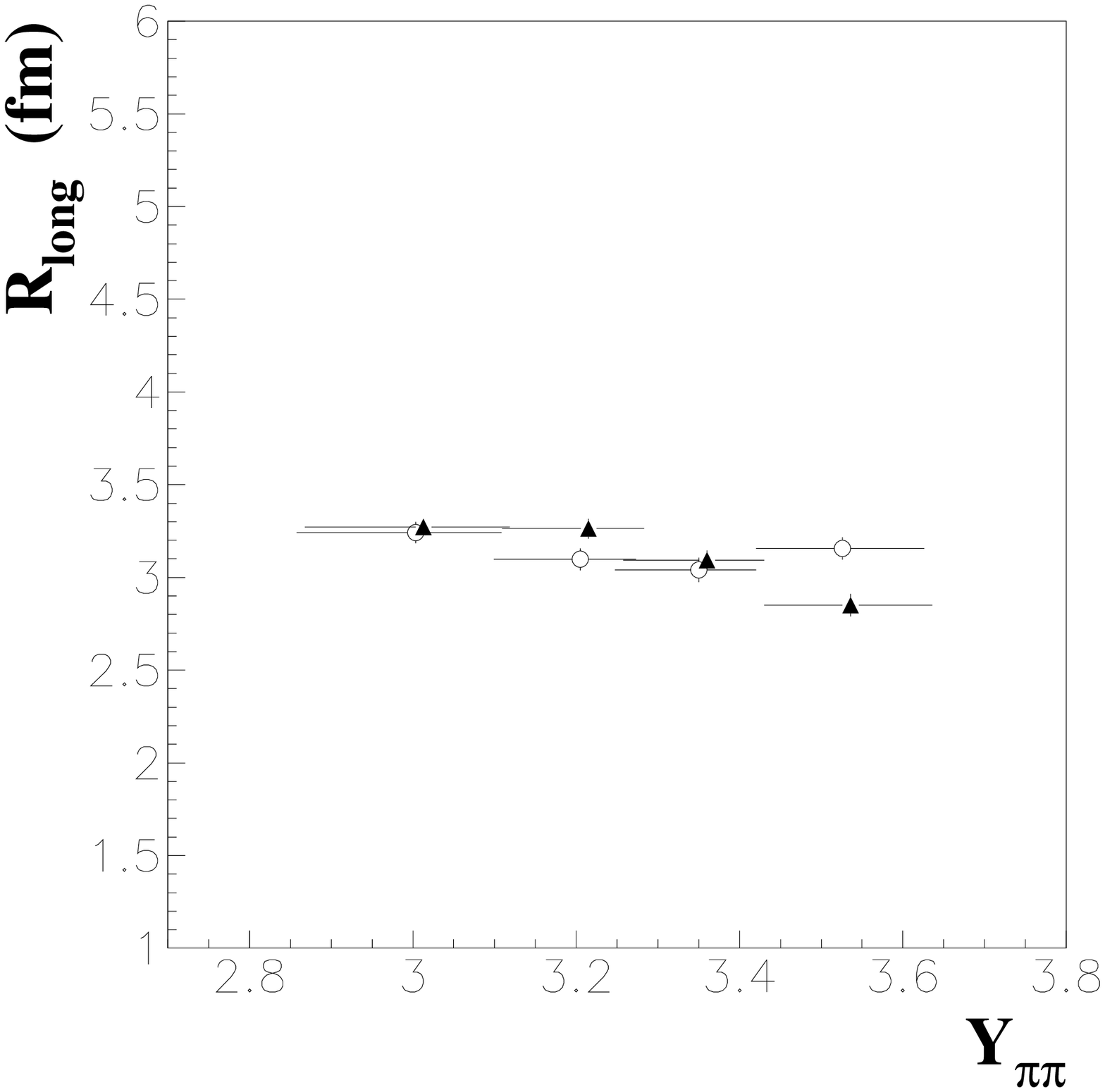}
\includegraphics{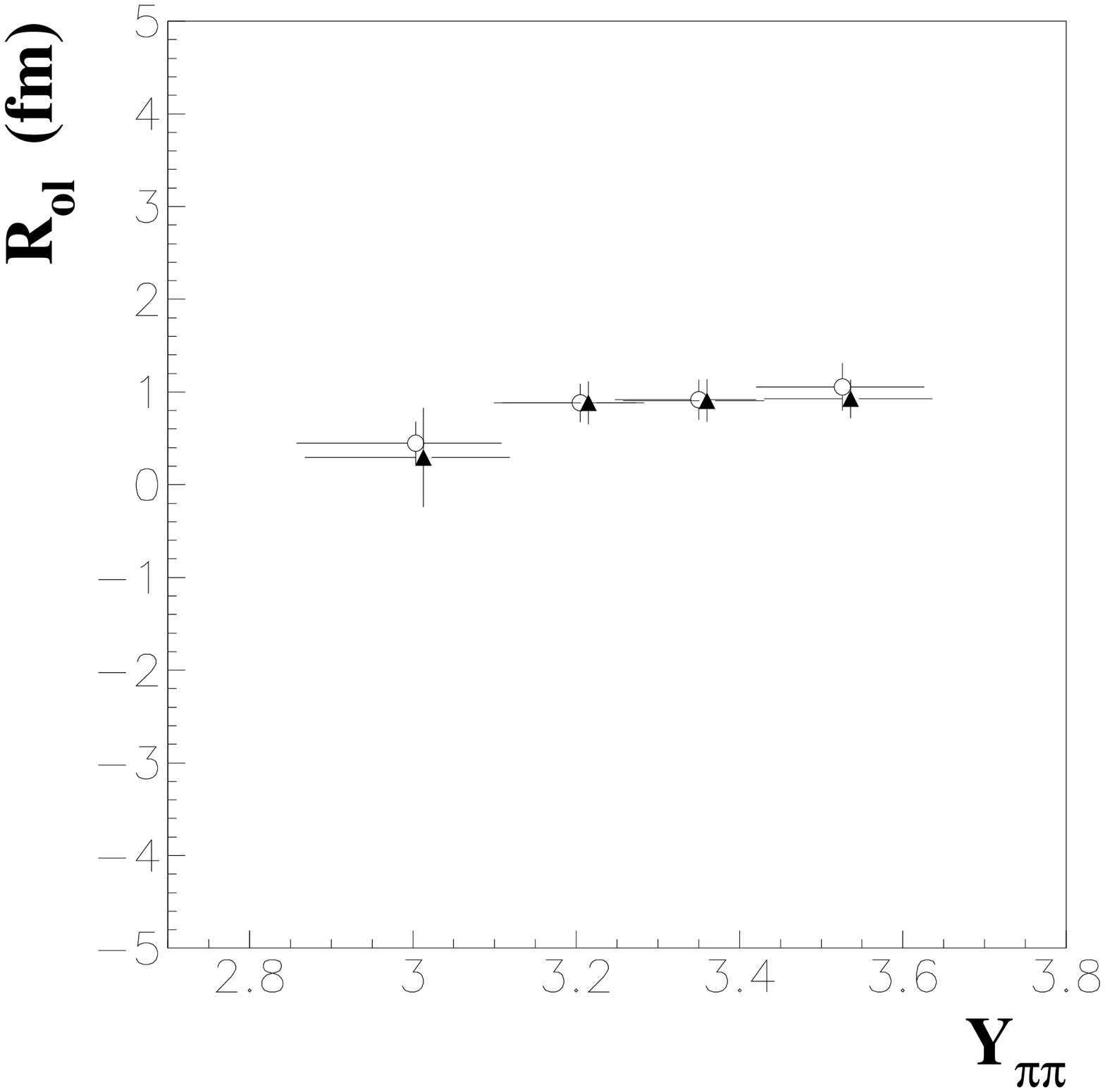}}\\
\caption{\rm Cartesian parameters as calculated directly (open circles) and 
from the YKP parametrization (black triangles) as a function of 
$Y_{\pi\pi}$\ in the interval  $0.23 < K_t < 0.76 \, {\rm GeV}/c$.}
\label{fig:comp_v}
\end{figure}
\par 
In fig.~\ref{fig:compil} we have compiled recent results 
from other 
CERN heavy-ion experiments, namely 
NA49~\cite{NA49,ref:Appel,comp:NA49}, 
WA98~\cite{comp:WA98} and NA44~\cite{comp:NA44},  
along with our results for the two most central collision classes 
(III and IV) merged together. 
\begin{figure}
  \centering
\resizebox{0.50\textwidth}{!}{%
\includegraphics{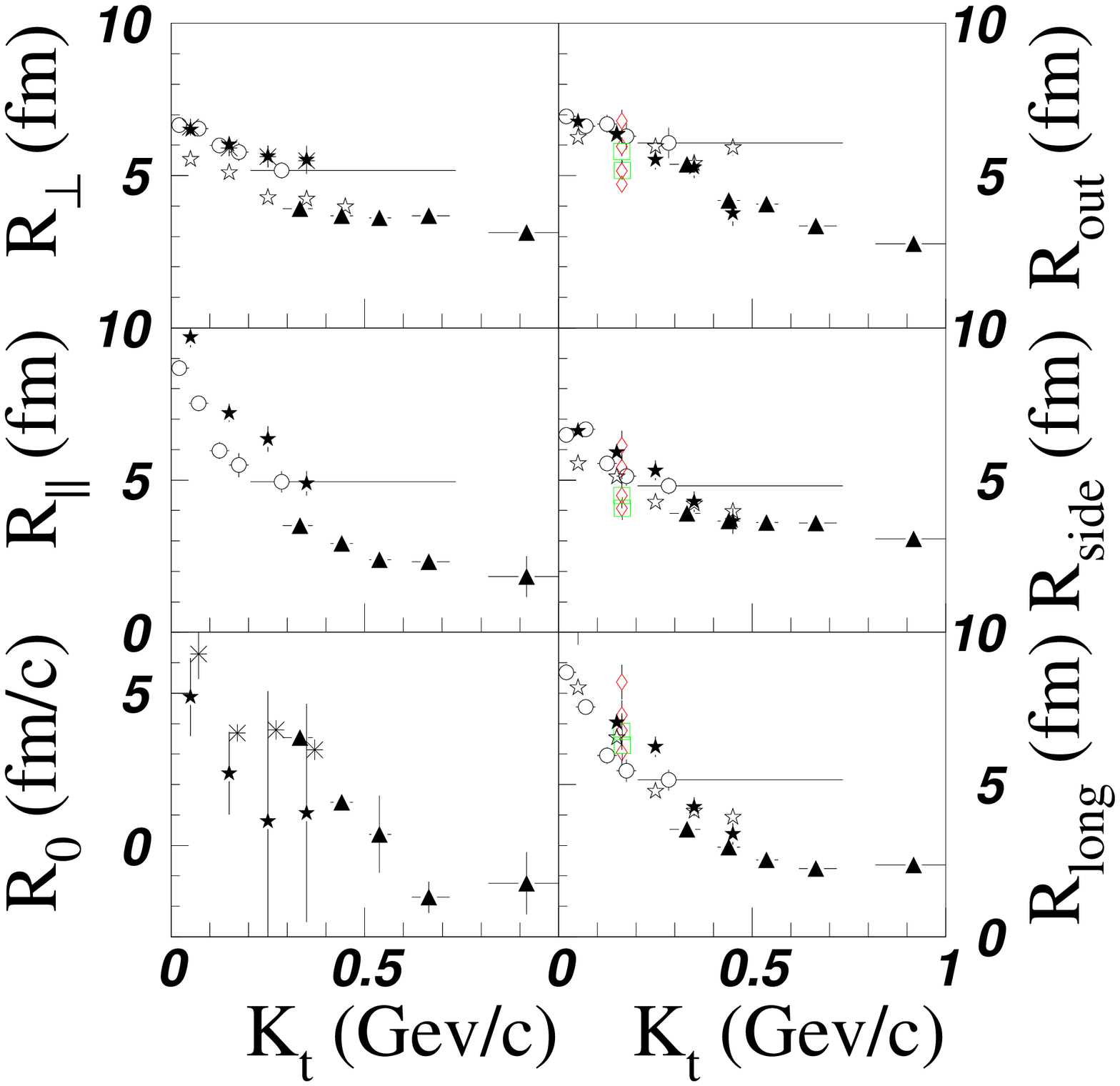}}\\
\vspace{0.1cm}
\resizebox{0.46\textwidth}{!}{%
\includegraphics{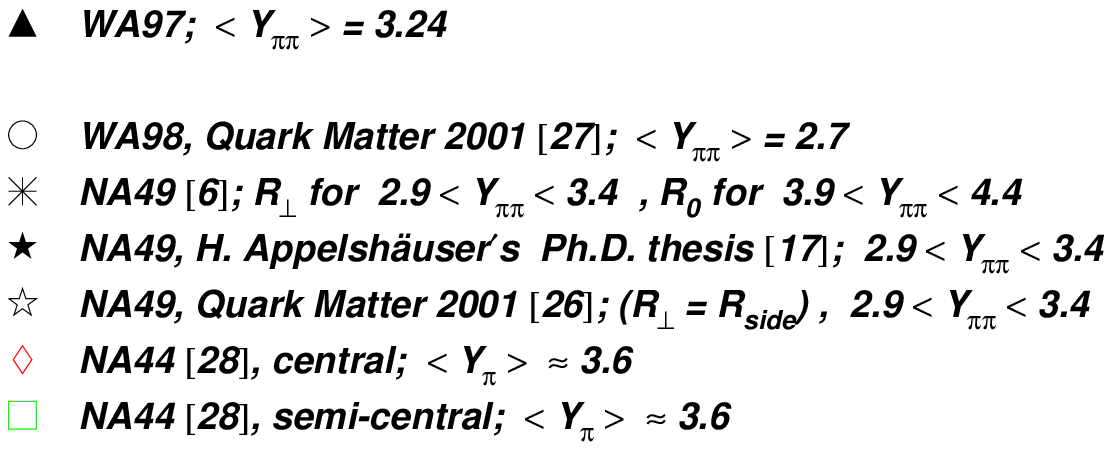}}
\caption{\rm Dependence of the HBT radii on the transverse 
 momentum of the pair for Pb-Pb collisions at SPS energy.}
\label{fig:compil}
\end{figure}
The overall agreement of our data with the other experiments 
is convincing both in the common $K_t$\ region and when  
extrapolating to small $K_t$. 
WA97 is the sole experiment that explores the high $K_t$\ 
region and our data suggest a flattening of the 
$R_{\perp}$, $R_{\parallel}$, $R_{s}$\ and $R_{l}$\ behaviour 
with $K_t$\ at high values. 
\subsection{The Yano-Koonin velocity versus $Y_{\pi\pi}$: longitudinal flow}
\label{vyk}

Fig.~\ref{fig:vyk} shows the YK rapidity 
$Y_{YK}=\frac{1}{2}\log{\frac{1+v_{yk}}{1-v_{yk}}} + Y_{\pi\pi}$, 
calculated from the 
fitted parameter $v_{yk}$, as a function of the pair rapidity $Y_{\pi\pi}$\ 
(both displayed in the laboratory system) separately for the four centrality 
classes and in the inclusive case (class ``ALL''). 
\begin{figure}
  \centering
  \resizebox{0.30\textwidth}{!}{%
  \includegraphics{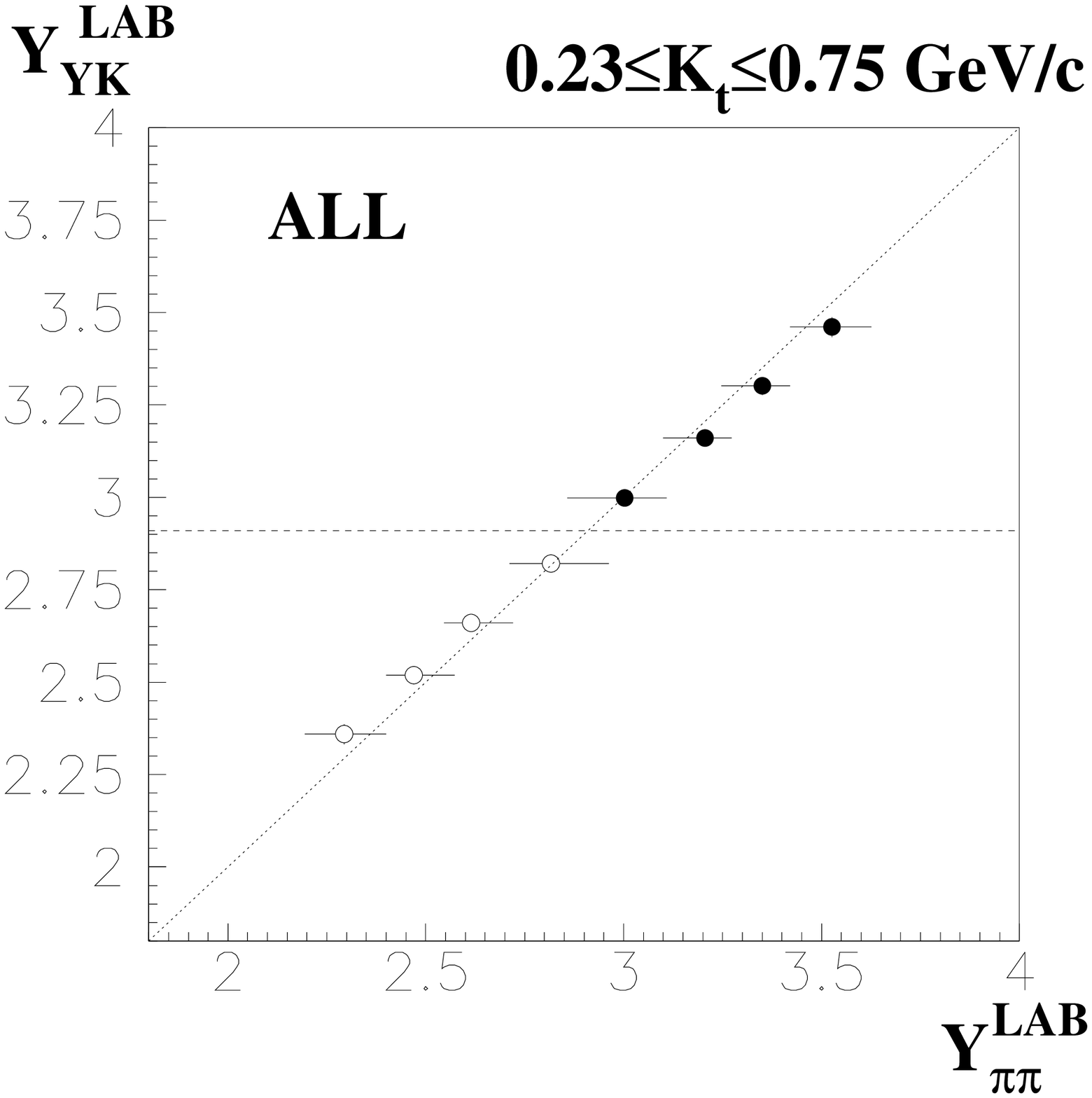}}\\
\resizebox{0.44\textwidth}{!}{%
\includegraphics{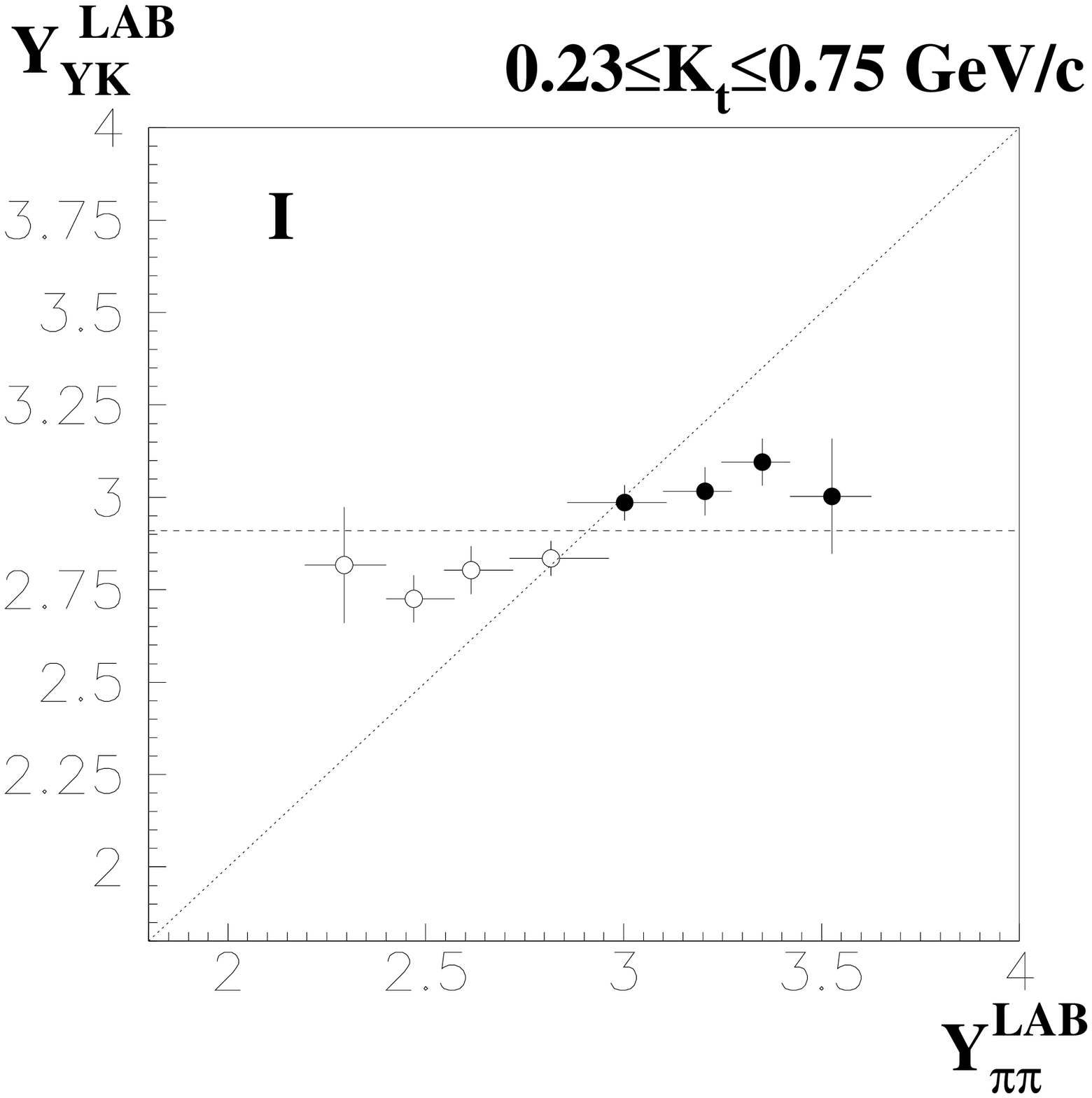}
\includegraphics{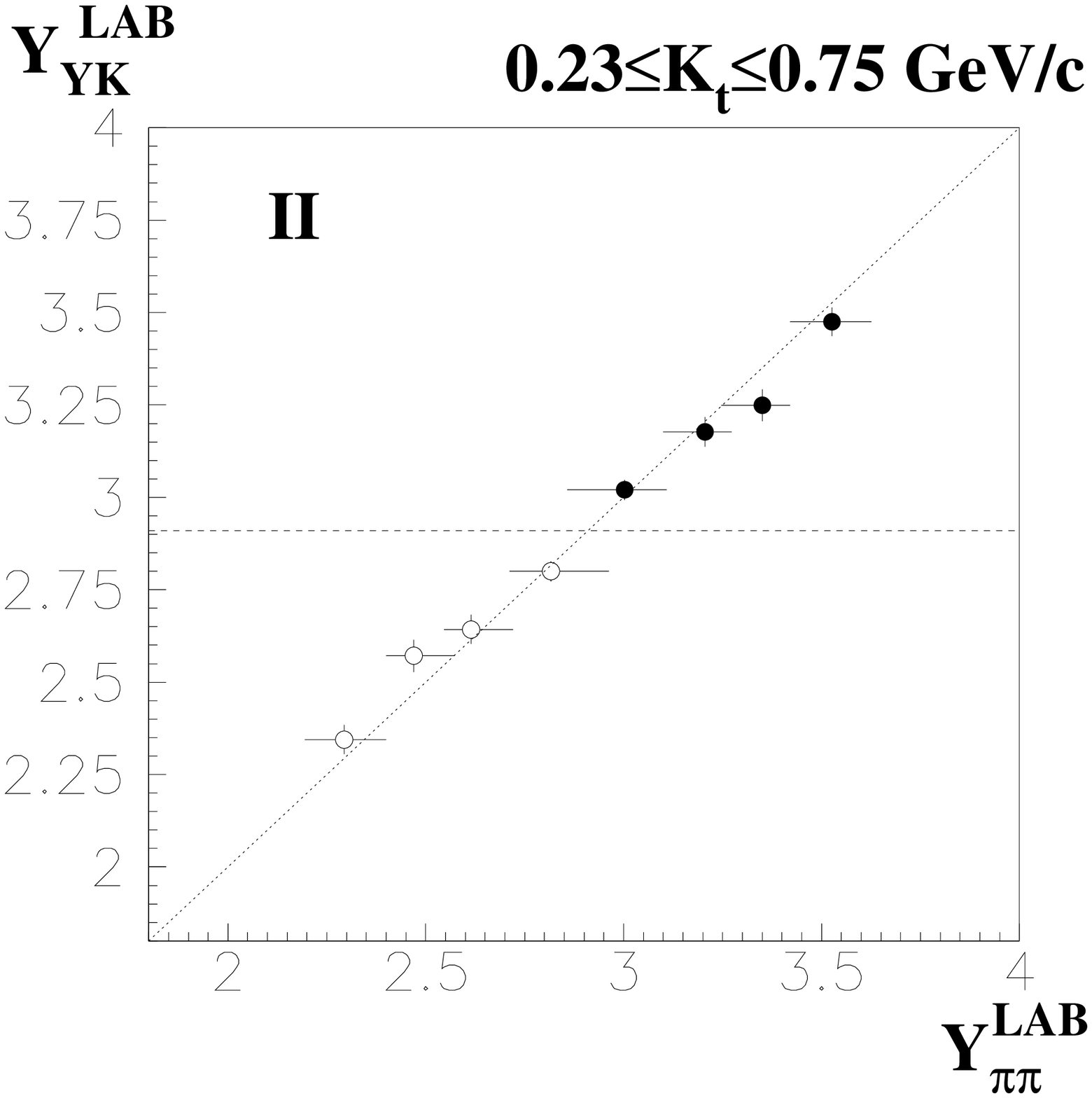}}\\
\resizebox{0.44\textwidth}{!}{%
\includegraphics{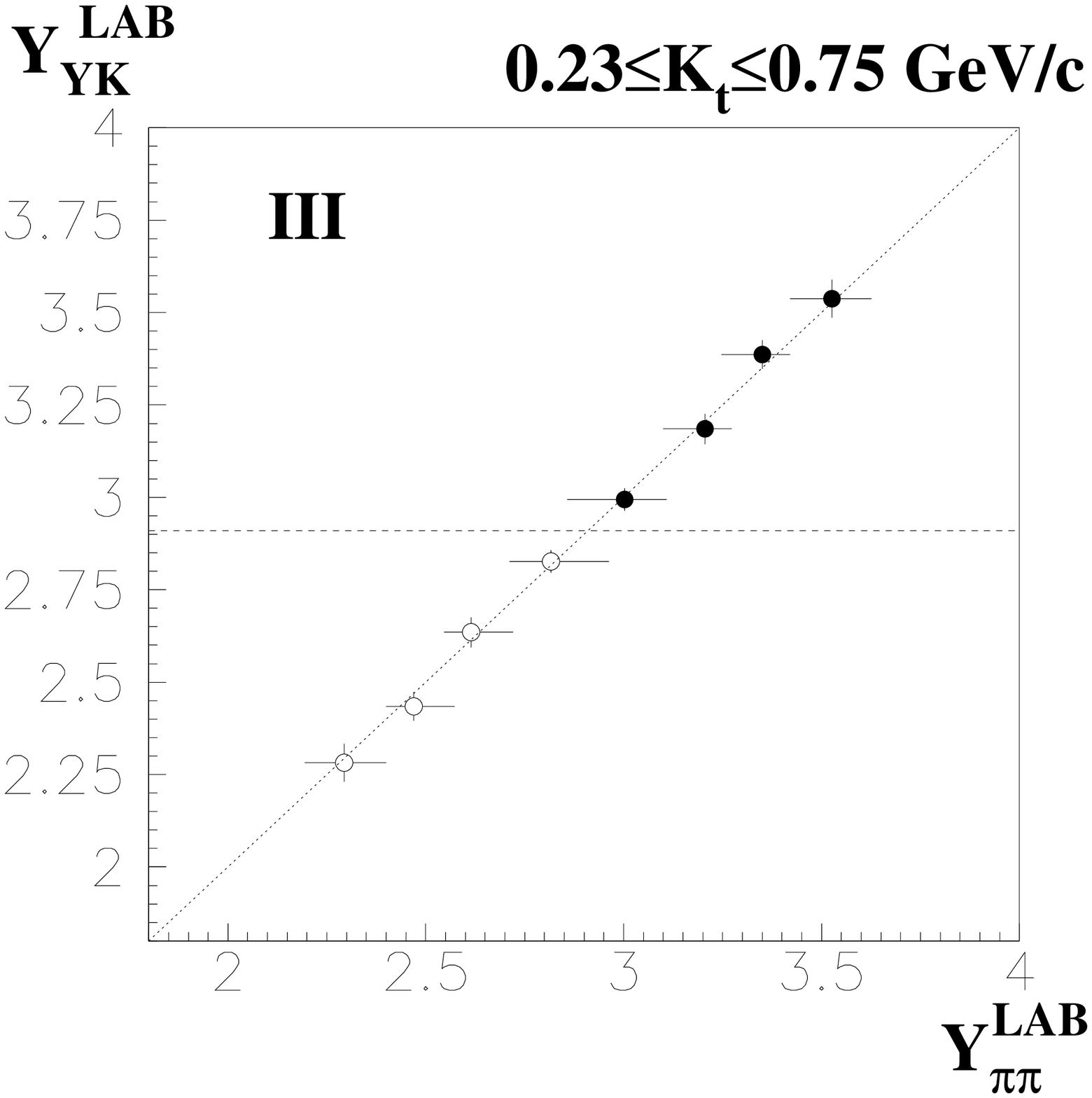}
\includegraphics{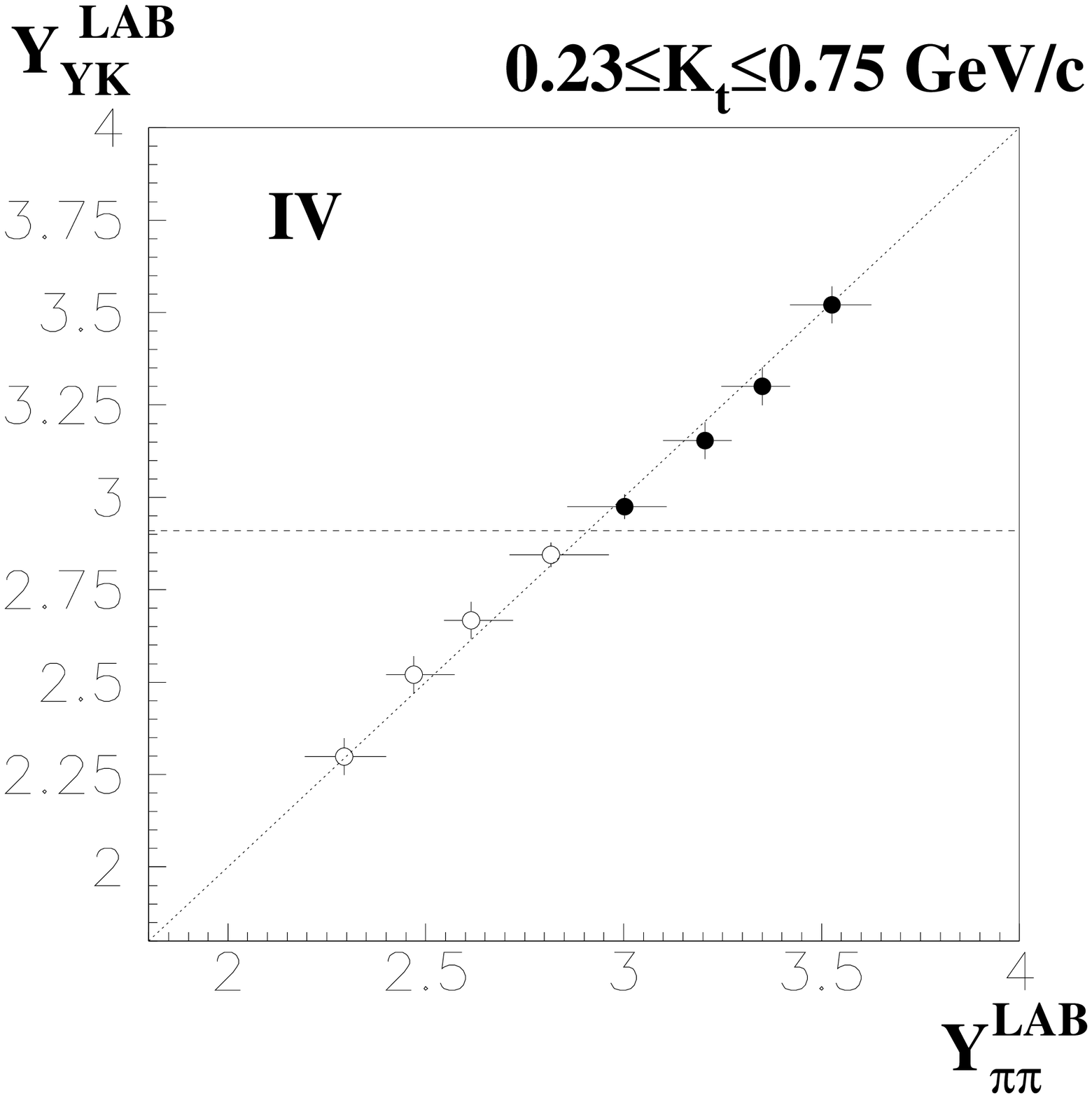}}
\caption{{\rm $Y_{\pi\pi}$\ dependence of the YK rapidity for the four 
 centrality classes and for the inclusive case (ALL). Class I corresponds 
 to the most peripheral collisions and class IV to the most central ones. 
 Filled circles are data, open circles are 
 data reflected about mid-rapidity.}}
\label{fig:vyk}
\end{figure}
The $v_{yk}$\ fit parameter is often determined 
with a larger relative error than $R_{\perp}$,
$R_{\parallel}$\ or $\lambda$ (the same is true for $R_0$); 
therefore four data sets (in the $K_t$\ variable) have been merged 
in order to obtain small errors on $v_{yk}$. 

As explained in ref.~\cite{RecProgr2} the dependence of the YK rapidity 
on the pair rapidity should directly measure the longitudinal expansion of 
the source and cleanly separate it from its transverse dynamics. 
In particular, for a non-expanding source $Y_{YK}$\ 
would be independent of $Y_{\pi\pi}$\ (horizontal lines in 
fig.~\ref{fig:vyk}), while for the case of 
a longitudinally expanding, boost-invariant source 
these variables 
would be completely correlated (diagonal lines in the same figure). 
It has been shown in ref.~\cite{Consistency2} 
that, as a consequence of the reduced thermal 
smearing at large $K_t$, the linear relation $Y_{YK}=Y_{\pi\pi}$\ 
is better satisfied the larger the transverse momentum of the 
particle pair. Hence this linear relation cannot be generally 
interpreted as evidence for a true boost-invariant longitudinal expansion.
The dependence of the YK velocity on $K_t$\ is expected however 
to be of minor importance. 

The results of fig.~\ref{fig:vyk}  thus would suggest 
that the source expands longitudinally 
in a nearly boost-invariant way for the three most central collision 
classes and with much less intensity for the most peripheral one.
We observe, however, an unexpectedly {\em strong} dependence of $v_{yk}$\ 
(hence of $Y_{YK}$) on the transverse pair momentum $K_t$, as discussed 
in the next section. 
\subsection{The Yano-Koonin velocity versus $K_t$:  
$v_{yk}$\ positive and close to $c$\ at high $K_t$} 
\label{vyk2}
Fig.~\ref{fig:vyk2} shows the $K_t$\ dependence of $v_{yk}$\ 
in the inclusive case of all
centrality classes merged together (class ``ALL''), and 
in the four centrality classes separately. 
In order to make a quantitative comparison of the behaviour of 
$v_{yk}$\ in the different classes, we have parametrized 
$v_{yk}$\ as a linear function of $K_t$: $v_{yk}=a + b K_t $. 
The slope $b$\ is found to be similar for all the four centrality 
classes with a value of about $1.5 \, ({\rm GeV}/c^2)^{-1}$, 
while the value of the intercept $a$\ is different for class I 
($a\approx -0.8 \,c$) with respect to classes II,III and 
IV ($a\approx -0.6 \,c$).  
The parametrization works 
satisfactorily up to $K_t \simeq 0.8 \, {\rm GeV}/c$, but 
\begin{figure}[hbt]
  \centering
\resizebox{0.24\textwidth}{!}{%
\includegraphics{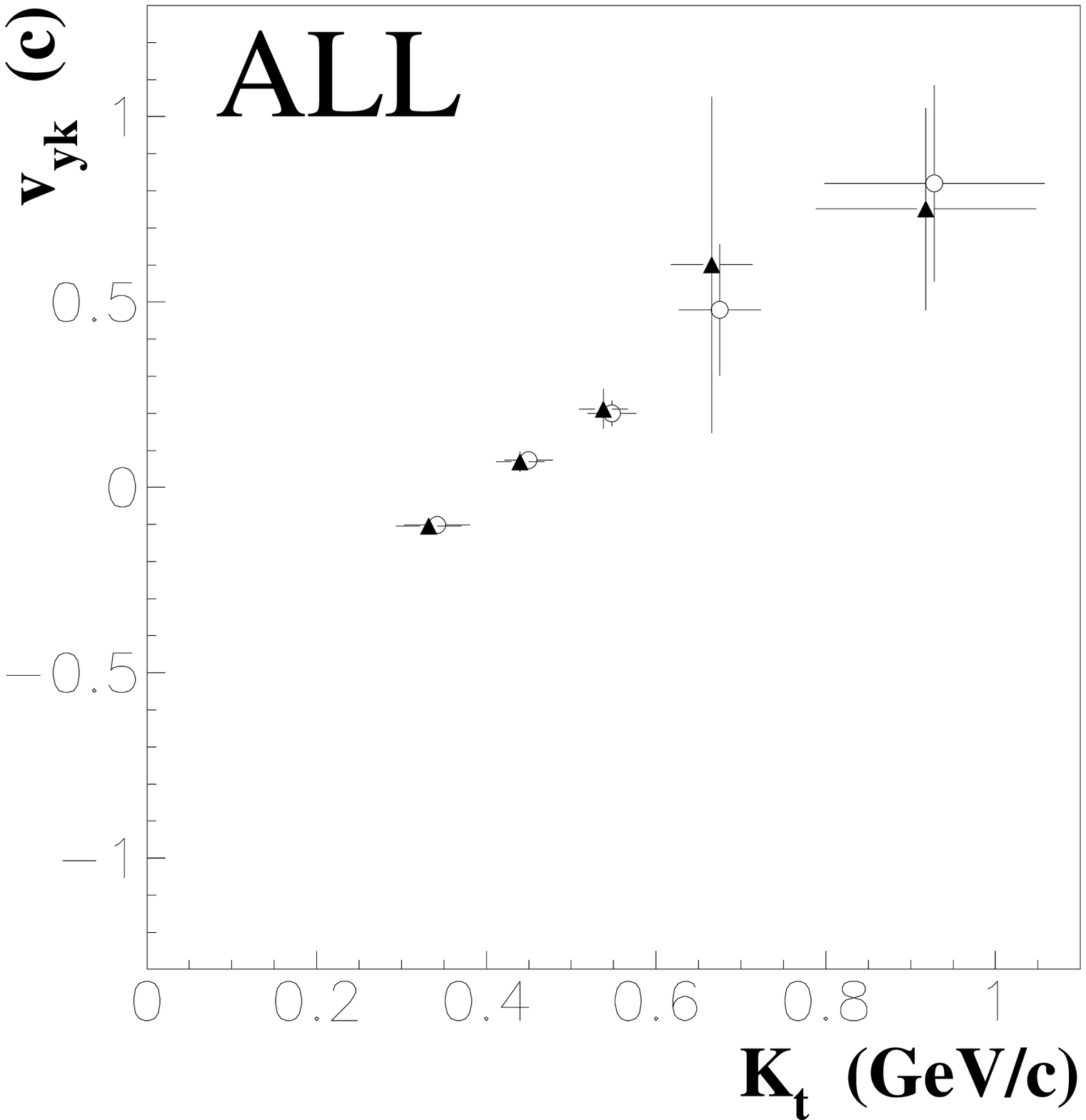}}\\
\resizebox{0.48\textwidth}{!}{%
\includegraphics{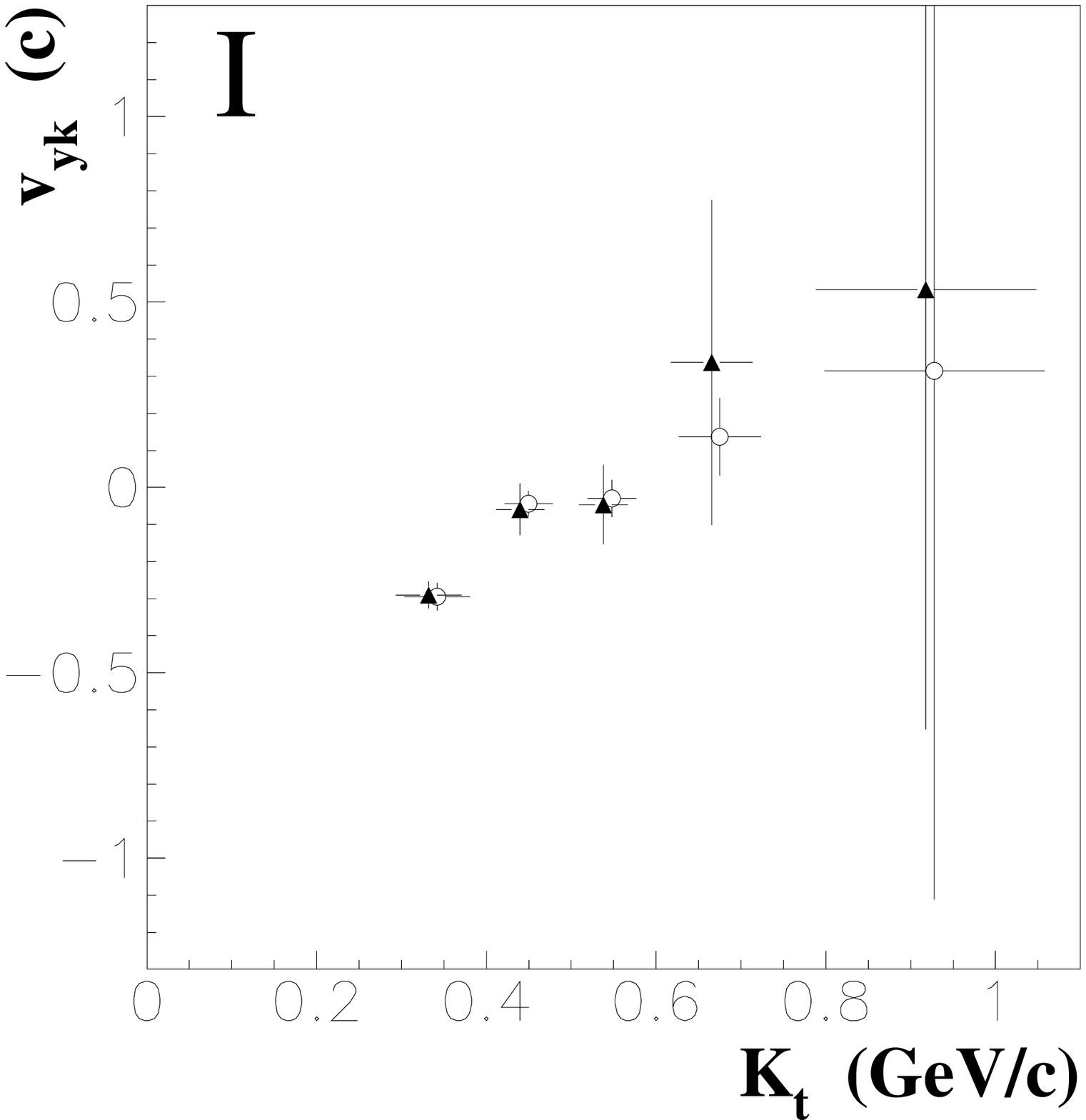}
\includegraphics{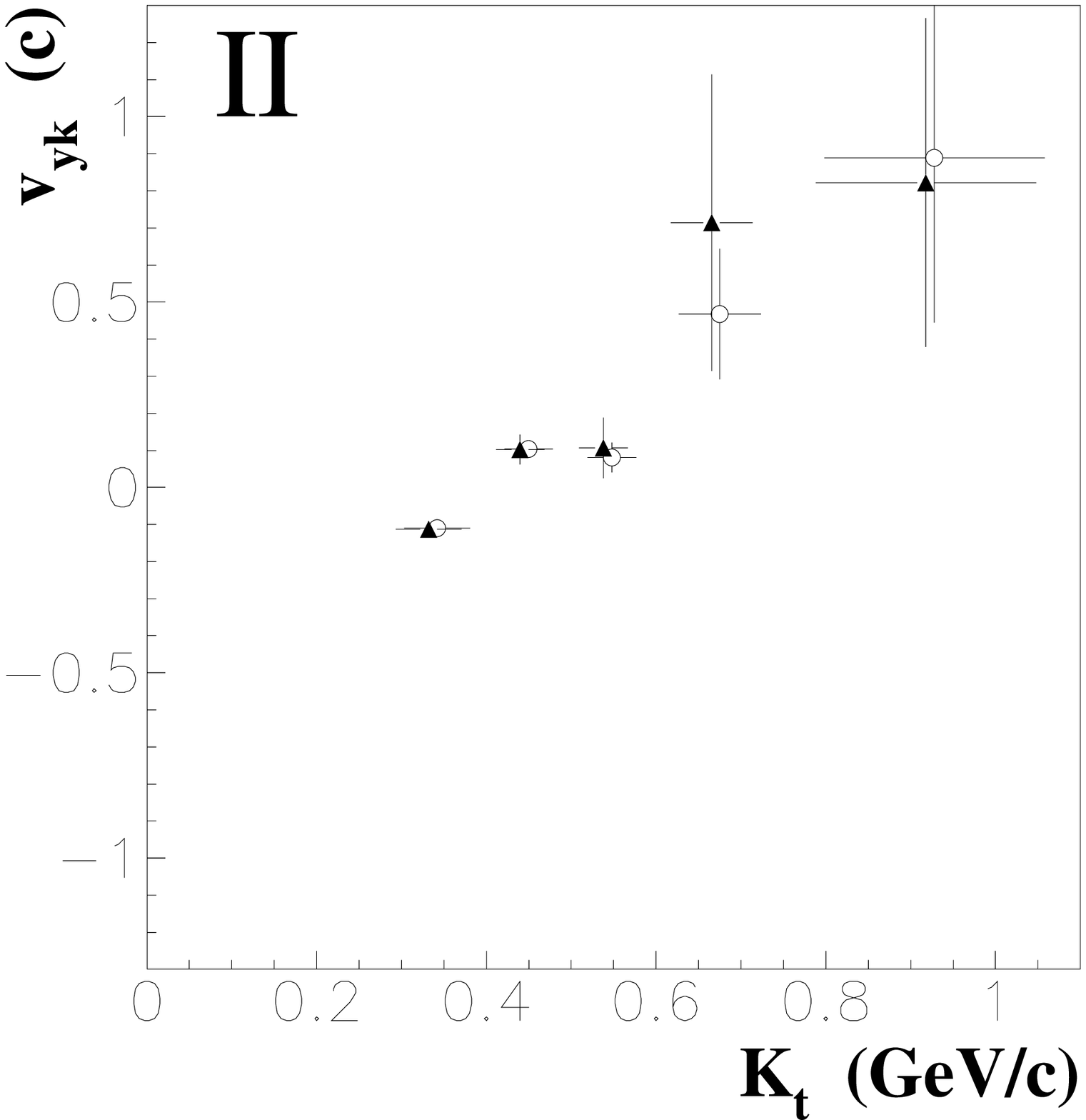}}\\
\resizebox{0.48\textwidth}{!}{%
\includegraphics{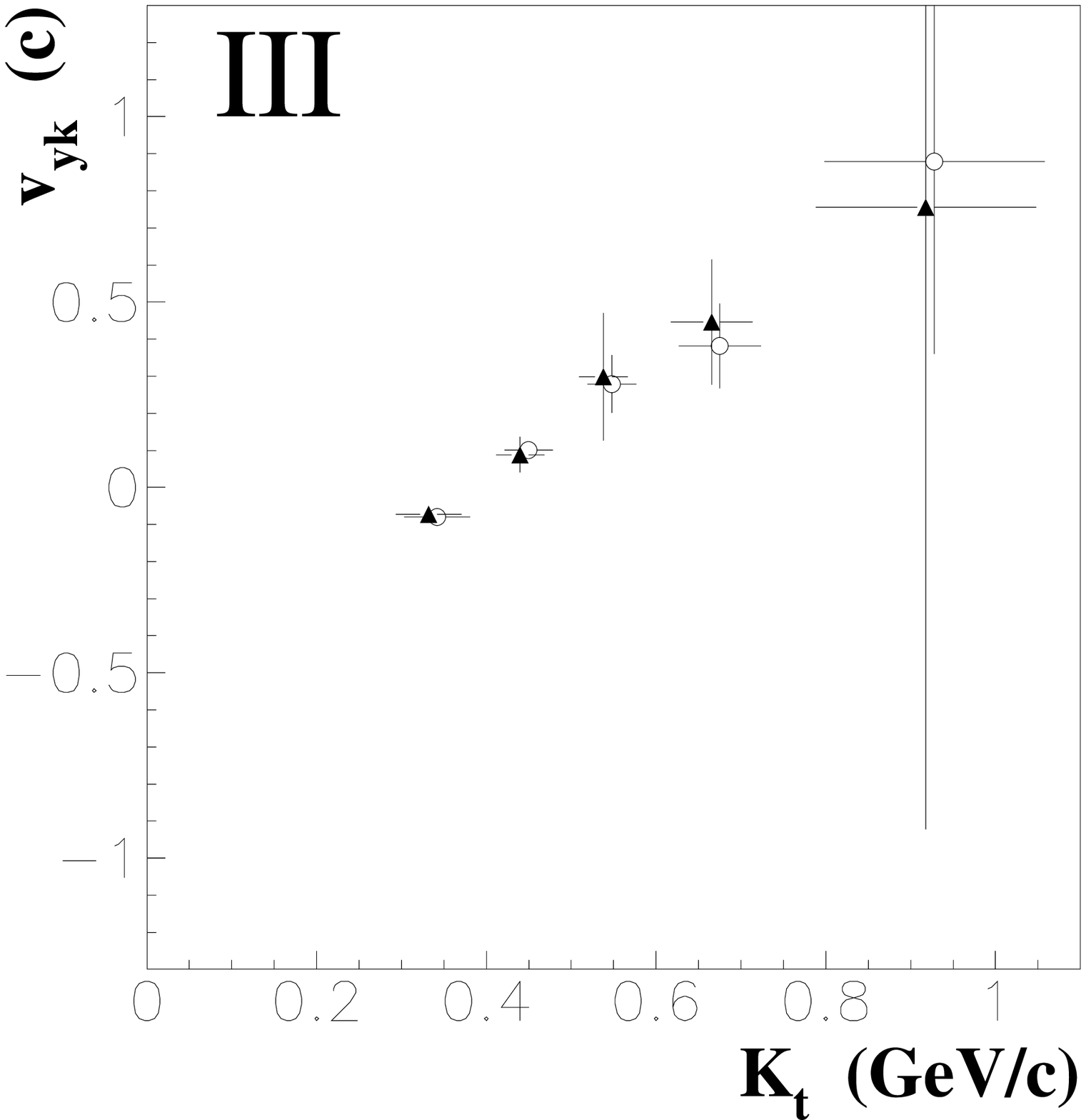}
\includegraphics{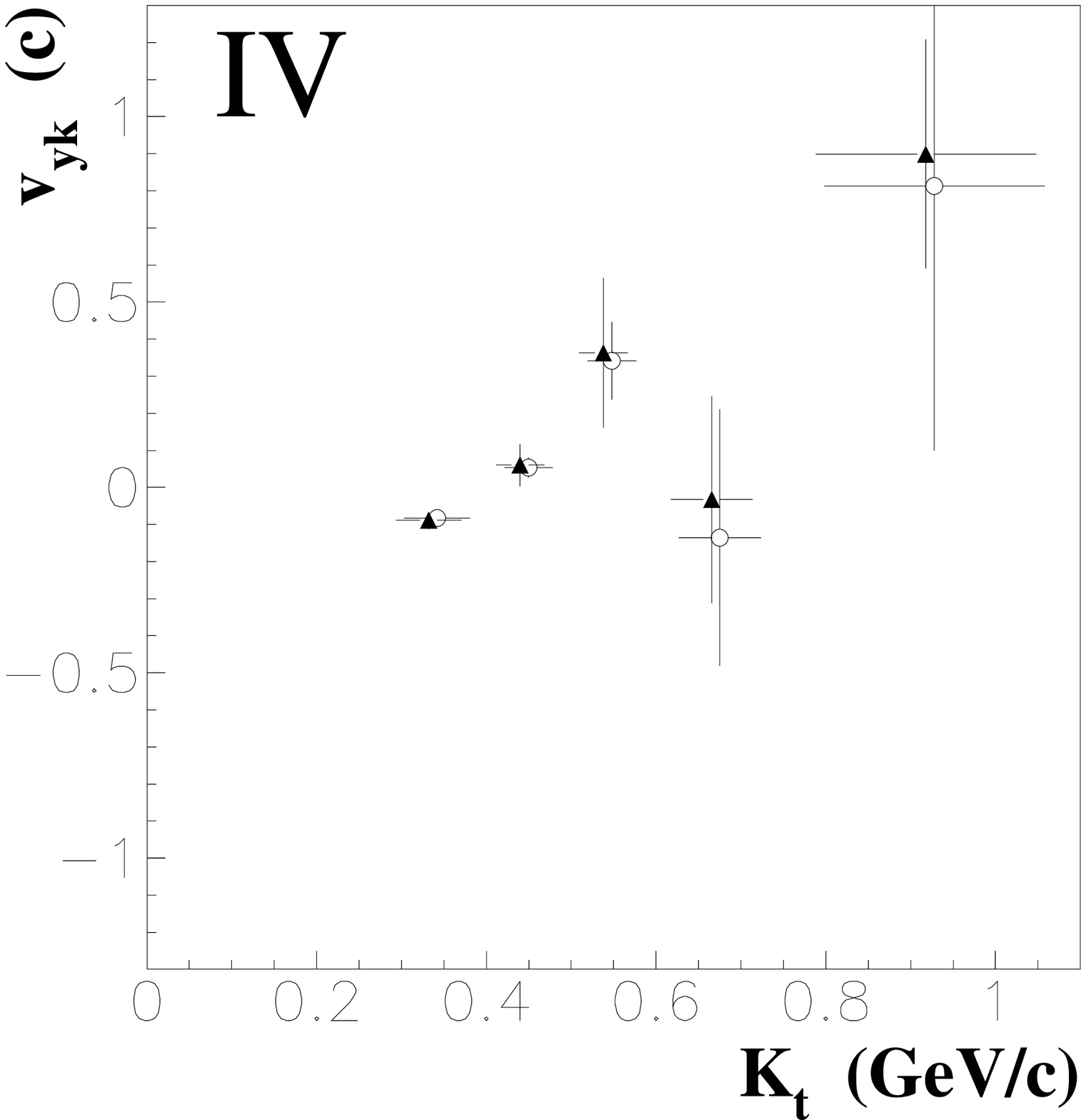}}
\caption{{\rm $K_{t}$\ dependence of the YK velocity in the four
 centrality classes and the inclusive (ALL) one. 
 Black triangles are $v_{yk}$'s measured 
 directly in the YKP parametrization, open circles are deduced 
 from Cartesian radii. 
}}
\label{fig:vyk2}
\end{figure}
the linear approximation becomes unphysical ($v_{yk}/c>1$) for 
$K_t > \frac{c-a}{b} \sim 1 \, {\rm GeV}/c$,  
therefore 
a saturation of $v_{yk}$\ up to a value close to the 
speed of lighthas to be supposed.
\par 
Positive $v_{yk}$\ in the LCMS means that the diagonal lines of 
fig.~\ref{fig:vyk} are crossed. 
Points in fig.~\ref{fig:vyk} are measured in the wide interval  
$ 0.23 < K_t < 0.75 \, {\rm GeV}/c$, the average $K_t$\ over this 
interval being however $\simeq 0.45 \, {\rm GeV}/c$. 
Since $v_{yk}$\ crosses the zero value for  
$K_t \sim 0.4 \, {\rm GeV}/c$, 
looking back at fig.~\ref{fig:vyk} we may claim that 
the experimental points are close to the $45^o$-degree line 
mainly because of their (mean) $K_t$\ values. 
We can conclude that the slopes of the plots 
$Y_{YK}^{SYS}$\ versus $Y_{\pi\pi}^{SYS}$\ 
($SYS$\ being either $LAB$\ or $CMS$), 
often encountered in literature, 
are influenced {\em strongly} by $K_t$. 
Therefore no conclusive predictions on the longitudinal expansion 
can be drawn by looking solely at the $Y_{\pi\pi}$\ dependence 
of the Yano-Koonin velocity. 
It is a fact, however, that 
we measure smaller $v_{yk}$\ for centrality class I: 
the $a$\ intercept has a larger negative value and the zero crossing 
occurs at higher $K_t$ for this class. Since the mean $K_t$\ 
of the pairs is found to be the same over the four centrality 
classes, therefore it can be deduced a less intense longitudinal 
expansion for class I.
\subsection{The $R_{\perp}$\ radius: transverse flow and temperature}
\label{Rt}

While longitudinal ``flow'' is not necessarily a signature for nuclear 
collectivity, because it may be due simply to incomplete stopping of 
the two colliding nuclei, {\em transverse} collective expansion flow 
can only be generated by the build up of a locally isotropic 
pressure component. 
The transverse radius parameter $R_{\perp}$\ is invariant under 
longitudinal boosts and thus is not affected at all by longitudinal 
expansion. It is expected to drop as a function of 
$M_t=\sqrt{m_{\pi}^2+K_t^2}$\ if the source expands in the 
transverse direction~\cite{RecProgr}.
\par
In ref.~\cite{RecProgr}, 
in the context of a source model characterized by 
longitudinal and transverse expansion, 
the following expression for $R_{\perp}$\ has been derived: 
\begin{equation}
R_{\perp}(K_t,Y_{\pi\pi})=
R_G \left[ 1+M_{t}\frac{\beta^2_{\perp}}{T}\cosh(Y_{YK}-Y_{\pi\pi})\right]
 ^{-\frac{1}{2}} 
\label{eq:Rt}
\end{equation} 
where $R_G$\ is equal to the transverse geometric 
(Gaussian) radius of the source times $1/ \sqrt{2}$, 
$T$\ is the freeze-out temperature and 
$\beta_{\perp}$\ the slope of the (linear) transverse flow velocity 
profile: $\beta_{\perp}(r)=\beta_{\perp}\frac{r}{R_G}$. 
A bidimensional fit of function~(\ref{eq:Rt}) to the experimental data 
has led to the determination of $R_G$\ and $\frac{\beta_{\perp}^2}{T}$.  
The fit results are summarized in table~3. \\ 
{\footnotesize {\bf Table 3.}  The transverse geometric parameter $R_G$\ 
 and the ratio $\beta^2_{\perp}/T$, $T$\ being the freeze-out temperature.}
\begin{center}
\begin{tabular}{||l|c|c|c||} \hline \hline
 Centrality & $ R_G \; ({\rm fm})$ &
$\frac{\beta_{\perp}^2}{T} \; ({\rm GeV}^{-1})$ &
$\chi^2/ndf $ \\ \hline
 ALL & $ 4.75 \pm 0.38 $   & $ 1.62 \pm 0.60 $ & $ 21.6/18 $ \\
  I  & $ 3.20 \pm 0.31 $   & $ 0.68 \pm 0.55 $ & $ 21.6/12 $ \\
 II  & $ 4.63 \pm 0.44 $   & $ 1.82 \pm 0.78 $ & $ 15.5/16 $ \\
 III & $ 5.03 \pm 0.56 $   & $ 1.97 \pm 0.92 $ & $ 20.6/17 $ \\
 IV  & $ 5.06 \pm 0.58 $   & $ 1.76 \pm 0.78 $ & $ 23.0/16 $ \\ \hline \hline
\end{tabular}
\end{center}
\vspace{0.6cm}

The $\vec{K}$\ dependence of $R_{\perp}$\ determines only the ratio 
$ \frac{\beta^2_{\perp}}{T} $, so that a full range for the 
$(\beta_{\perp},T)$\ 
doublet is still conceivable. WA97 has measured the inverse $m_t$\ slope 
parameters $T_{eff}$\ as a function of centrality~\cite{TranSlope}. 
This slope can be thought as the inverse of a blue-shifted temperature 
$T_{eff}=T\sqrt{\frac{1+\beta_{\perp}}{1-\beta_{\perp}}}$. 
This relation provides a second 
constraint on the $(\beta_{\perp},T)$\ doublet range. 
Combining the {\em two}-particle correlation measurements 
with the {\em single} particle transverse mass spectra, 
disentangles temperature $T$\ and transverse flow $\beta_{\perp}$. 
The allowed regions in the 
$(T,\beta_{\perp})$\ parameter space, 
as determined by the combination of these two measurements, 
are shown in fig.~\ref{fig:T-betat}, and the disentangled values 
for $T$\ and $\beta_{\perp}$\ are reported in table~4. \\ 
\vspace{0.3cm} \\
{\footnotesize {\bf Table 4.} The freeze-out temperature $T$\ and the 
intensity of the transverse flow $\beta_{\perp}$.}
\begin{center}
\begin{tabular}{||l|c|c||} \hline \hline
 Centrality & $ T ({\rm MeV})$ & $\beta_{\perp}$ \\ \hline
 ALL & $ 123^{+12}_{-10} $  & $ 0.44^{+0.06}_{-0.08} $ \\
  I  & $ 140^{+26}_{-13} $  & $ 0.30^{+0.09}_{-0.16} $ \\
 II  & $ 121^{+15}_{-11} $  & $ 0.47^{+0.07}_{-0.10} $ \\
 III & $ 117^{+16}_{-11} $  & $ 0.48^{+0.08}_{-0.11} $ \\
 IV  & $ 120^{+15}_{-11} $  & $ 0.46^{+0.07}_{-0.10} $ \\  \hline \hline
\end{tabular}
\label{tab:T-Betat}
\end{center}
\vspace{0.6cm}

Postponing to the next section a global discussion about these results, 
we wish here to emphasize that class I features a smaller 
transverse freeze-out radius, a lower transverse expansion velocity 
and a higher temperature. 
This suggests an expansion on a smaller scale; the higher temperature 
at the freeze-out may be interpreted as the remnant of 
an earlier decoupling of the expanding system. 

\begin{figure}[t]
\centering
\resizebox{0.44\textwidth}{!}{%
\includegraphics{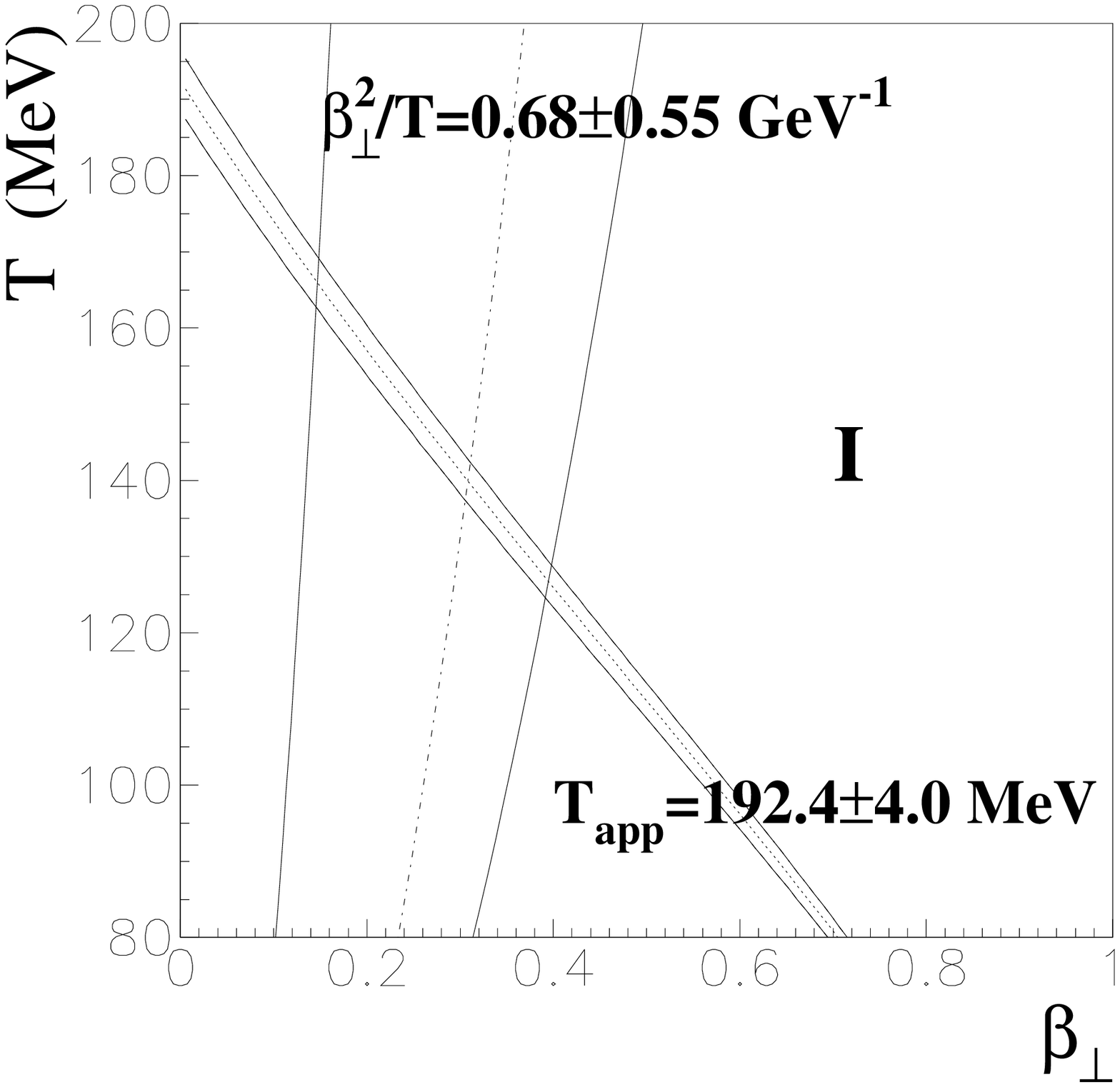}
\includegraphics{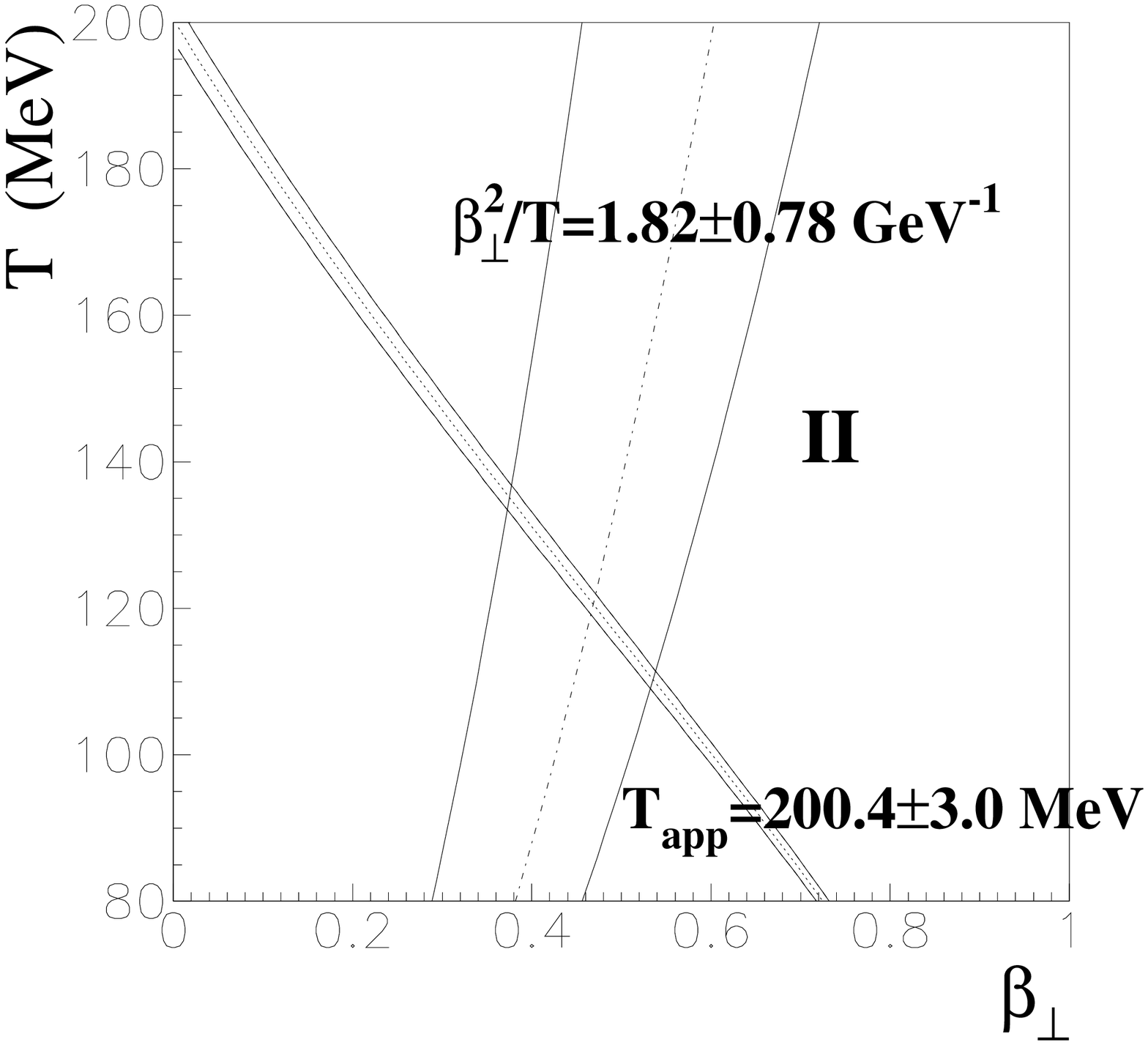}}\\
\resizebox{0.44\textwidth}{!}{%
\includegraphics{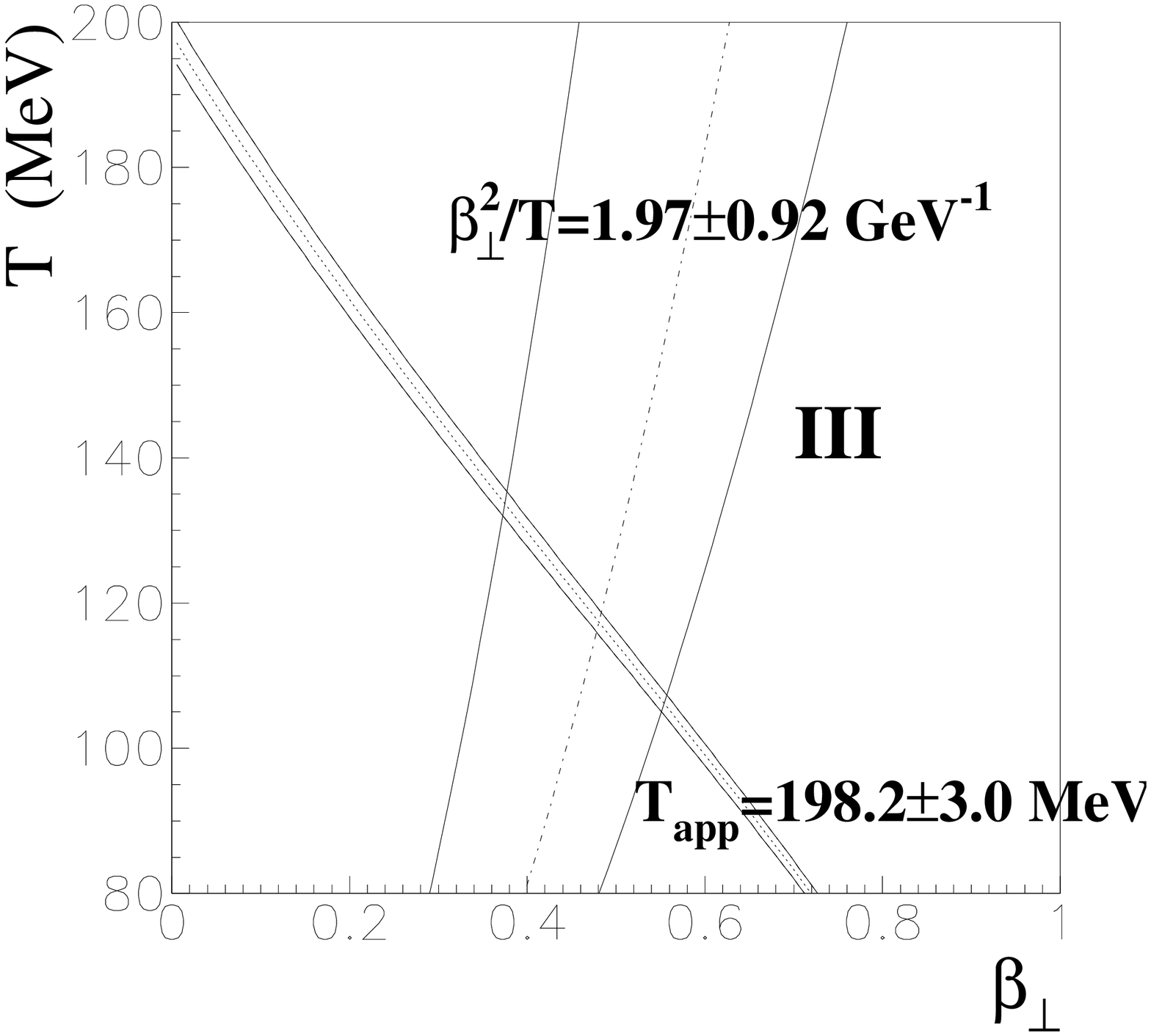}
\includegraphics{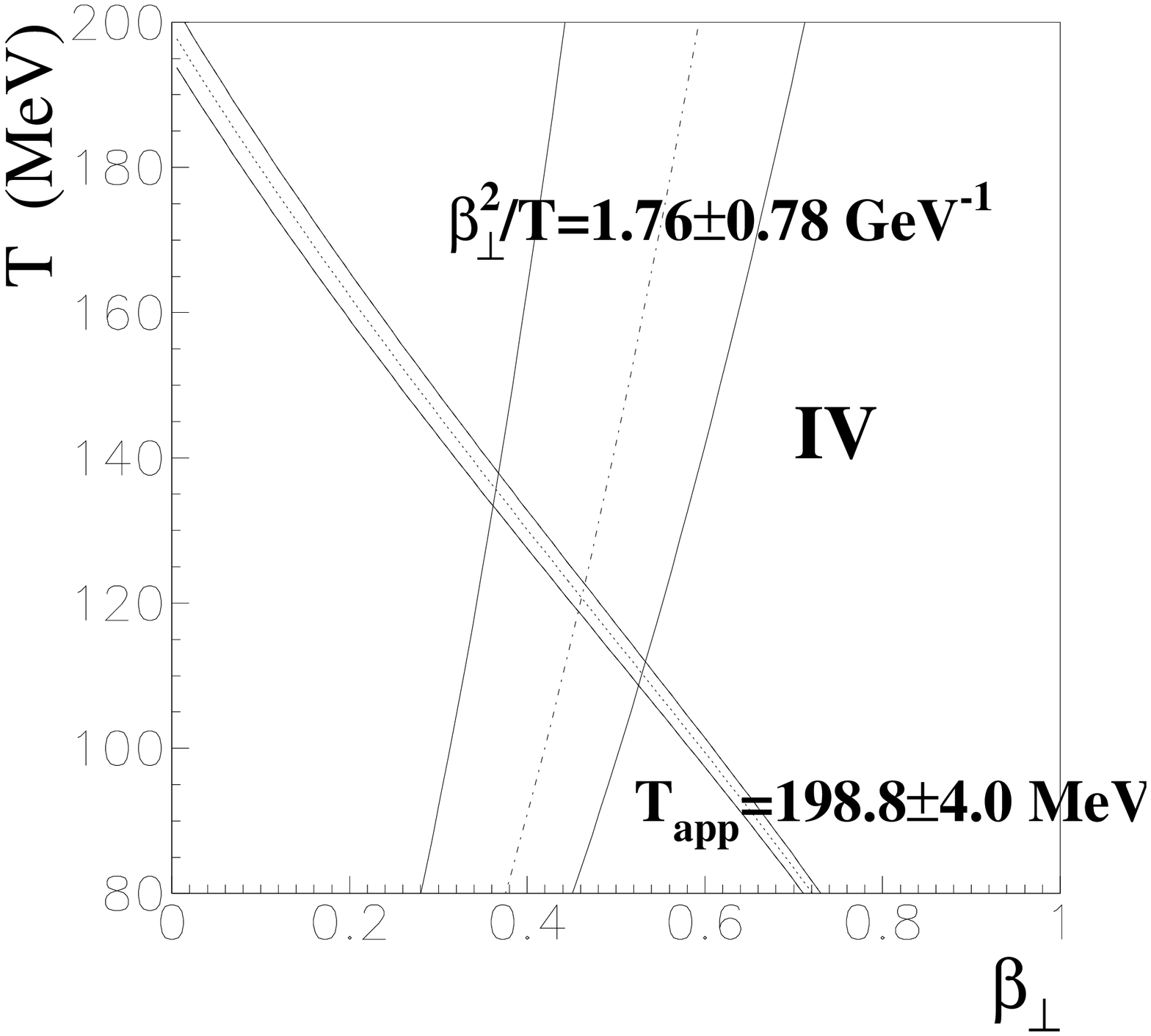}}\\
\caption{{\rm Allowed regions 
in the freeze-out temperature vs. transverse flow 
velocity space 
(see text). Bands are drawn at $\pm \sigma$\ around fitted values.}}
\label{fig:T-betat}
\end{figure}
\subsection{The $R_{\parallel}$\ radius: average freeze-out proper time}
\label{Rp}
In the same model of ref.~\cite{RecProgr}, the dependence of 
$R_{\parallel}$\ on $(K_t,Y_{\pi\pi})$\ determines the average 
freeze-out proper time $\tau_0$\ via the formula:
\begin{equation}
R_{\parallel}= \tau_o[\frac{M_t}{T}\cosh(Y_{YK}-Y_{\pi\pi}) - 
  \cosh^{-2}(Y_{YK}-Y_{\pi\pi}) +1/(\Delta\eta)^2]^{-\frac{1}{2}}
\label{eq:Rperp}
\end{equation}
The parameter $\Delta\eta$\ is fixed by the extension of the 
single particle rapidity distribution~\footnote{ 
$\Delta\eta$\ is related to the width of the rapidity distribution 
$\Delta y$\ by $(\Delta y)^2=(\Delta\eta)^2+\frac{T}{m_T}$.
}
; it has been measured by the large acceptance NA49 
experiment to be $\Delta\eta=1.3 \pm 0.1$~\cite{NA49Deltaeta}. 
The results of the bidimensional fits of function~(\ref{eq:Rperp}) 
to the experimental data are presented in table~5.\\ 
\vspace{0.3cm} \\
{\footnotesize {\bf Table 5.}  Average freeze-out proper time 
 $\tau_0$\ of the source.}
\begin{center}
\begin{tabular}{||l|c|c||} \hline \hline
 Centrality & $\tau_0 \; ({\rm fm}/c)$ &
$\chi^2/ndf $ \\ \hline
 ALL & $ 5.35 \pm 0.20 $   & $ 8.6/19  $ \\
  I  & $ 3.66 \pm 0.22 $   & $ 24.3/13 $ \\
 II  & $ 5.08 \pm 0.23 $   & $ 16.5/17 $ \\
 III & $ 5.56 \pm 0.23 $   & $ 20.9/19 $ \\
 IV  & $ 5.58 \pm 0.24 $   & $ 9.6/17  $ \\  \hline \hline
\end{tabular}
\end{center}
\vspace{0.6cm}

The centrality classes III and IV show a similar behaviour, 
being characterized by a common mean freeze-out (proper) time 
$\tau_0 \simeq 5.6 \,{\rm fm}/c$, class II presents a sligthly shorter 
$\tau_0$\ of about  $5\, {\rm fm}/c$. All these values are much larger 
than the $3.7 \,{\rm fm}/c$\ measured for class I. 
\subsection{The $R_0^2$\ parameter: mean duration of the emission}
\label{R0}
The YKP parameter $R_0$ is related to the mean emission duration 
$\Delta\tau$\ of the source~\cite{RefDeltatau}:
\begin{equation}
 \Delta\tau \, \simeq R_0  \quad {\rm at \; high}\;  K_t
\label{eq:Deltatau}
\end{equation}
As one can see for the inclusive case of {\em ALL} centralities shown 
in fig.~\ref{fig:R0-Y},  the $R_0$\ radius does not 
present a significant dependence on the pair rapidity $Y_{\pi\pi}$. 
Therefore we have merged all the four $Y_{\pi\pi}$\ bins with the 
aim of reducing the statistical fluctuations on the 
extracted $R_0$\ when the  
four centrality classes are investigated separately. 
\begin{figure}[hbt]
\centering
   \resizebox{0.43\textwidth}{!}{%
\includegraphics{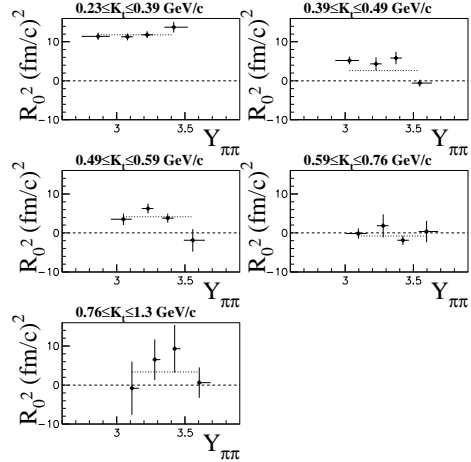}}
\caption{{\rm The $R_0^2$\ parameter as a function of $Y_{\pi\pi}$\  
 in five successive $K_t$\ intervals. The dotted horizontal lines through 
 the points are the best-fit results with a constant function.
  The showed data refer to the case of ``ALL'' centralities.}}
\label{fig:R0-Y}
\end{figure}
The $R_0^2$\ fit results for these data sets are displayed in 
fig.~\ref{fig:R0-Kt} as a function of $K_t$. 
\begin{figure}
\centering
\resizebox{0.45\textwidth}{!}{%
\includegraphics{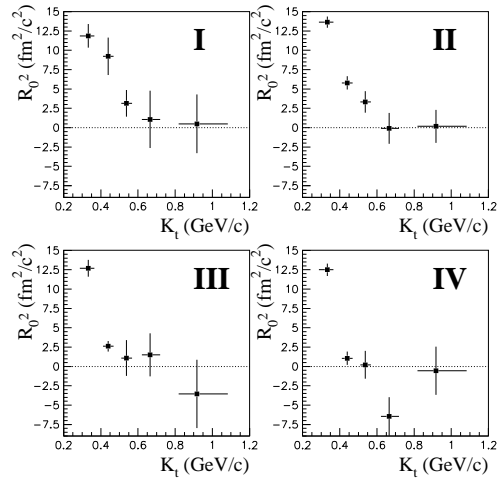}}
\caption{{\rm Dependence of $R_0^2$\ on $K_t$\ in our full rapidity 
interval, for the different classes of collision centrality.}}  
\label{fig:R0-Kt}
\end{figure}
\par
Negative values of the $R_0^2$\ parameter can be expected 
from an opaque source even at $K_{\perp}=0$~\cite{37-38}. 
In an opaque source the freeze-out occurs in the form of 
surface emission, as 
in the case of photons radiated by the sun. Fig.~\ref{fig:R0-Kt} 
suggests that all centrality classes are compatible with a transparent 
source that suddenly freezes out ($\Delta\tau \, \sim \, 0$). 
\section{Discussion}
\label{sec_disc}
The study of the $\vec{K}$--dependence of YKP parameters, combined 
with single-particle transverse-mass spectra, has determined several 
features of the source: its freeze-out temperature $T$, its 
transverse geometric (Gaussian) radius $\sqrt{2}R_G$, its average 
freeze-out proper time $\tau_0$, the mean proper time duration of 
particle emission $\Delta\tau$, the strength of the transverse flow 
$\beta_{\perp}$, some hints on the longitudinal expansion. 
In table~6 we summarize the main results of this study. 
These data provide implicitly a dynamical picture of the collision 
process. \\ 
\vspace{0.3cm} \\
{\footnotesize {\bf Table~6.} 
Source parameters for the four centrality classes.}
\begin{center}
{\footnotesize
\begin{tabular}
{||p{0.5in}|c|c|c|c||} \hline \hline
   & IV & III & II & I \\ \hline
 $R_G ({\rm fm})$  & $5.1 \pm 0.6$ & $5.0 \pm 0.6$ & $4.6 \pm 0.4$
                   & $3.2 \pm 0.3$   \\ \hline
 $T ({\rm MeV})$   & $120^{+15}_{-11}$ & $117^{+16}_{-11}$ &
                     $121^{+15}_{-11}$ & $140^{+26}_{-13}$  \\ \hline
 $\beta_{\perp} (c)$  & $0.46^{+0.07}_{-0.10}$ & $0.48^{+0.08}_{-0.11}$ &
                     $0.47^{+0.07}_{-0.10}$ & $0.30^{+0.09}_{-0.16}$ 
                     \\ \hline
 $\tau_0 ({\rm fm}/c)$& $5.6\pm 0.2$ & $5.6 \pm 0.2$ &
                        $5.1\pm 0.2$ & $3.7 \pm 0.2$   \\ \hline
 $\Delta\tau({\rm fm}/c)$ &
 \multicolumn{4}{c||}{$\approx 0$\ at high $K_t$} \\ \hline
 Long. expans.
  & \multicolumn{3}{c|}{fast} & slow \\ \hline \hline
\end{tabular}}
\end{center}
\vspace{0.6cm}
\par
With the aim of seeing if this picture is self-consistent, 
we can compare the two-dimensional rms width 
$R_{rms}^{freezeout}$\ $=  \sqrt{2} R_G$\ with the two-dimensional rms 
width of a cold lead nucleus 
\begin{equation}
 R_{rms}^{Pb}=\sqrt{\frac{3}{5}}1.2{\rm A}^{\frac{1}{3}}
 \simeq 4.5 \, {\rm fm.}
\label{ciccio}
\end{equation} 
For most central collisions (class III and IV), the system expands by 
a factor $\sim 1.6$ or, equivalently, by $2.65 \, {\rm fm}$\ in the 
transverse direction. If the transverse flow velocity is equal 
to $\beta_{\perp} \simeq 0.47 \, c$\ during the whole expansion, 
in a time of $\tau_0\simeq 5.6 \, {\rm fm}/c$\ nuclear matter 
would travel over $\tau_0\beta_{\perp}\simeq 2.6 \, {\rm fm}$\ 
in the transverse direction. This is consistent with the previous estimate 
from the difference $ R_{rms}^{freezeout} - R_{rms}^{Pb} $. 
Therefore the HBT description is internally consistent.
\par
For less central collisions (classes I,II) the initial overlap 
of colliding nuclei is expected to be smaller than that 
given by eq.(~\ref{ciccio}). A realistic estimation of 
such an overlap is not straightforward.  However, a backward reconstruction 
from the data  is still feasible. 
If the freeze-out transverse two-dimensional rms width is 
$4.5 \, {\rm fm}$\ ($6.5 \, {\rm fm}$) for class I (class II), 
accounting for an overall expansion of 
$\tau_0 \beta_{\perp}=1.1 \, {\rm fm}$\ ($2.4 \, {\rm fm} $), we  
may estimate the initial transverse rms width to be $3.4 \, {\rm fm}$\ 
($4.1 \, {\rm fm}$). 
\par
No quantitative description of the longitudinal expansion can be done. 
We can qualitatively say that the longitudinal expansion becomes stronger 
the more central the collisions; additionally our data would suggest 
a sudden step 
passing from centrality class I to class II 
(see fig.~\ref{fig:vyk}). 
\par
Concerning the collision time development, we have argued that the pion 
emission is a sudden process ($\Delta\tau \sim 0$), starting after 
$\sim 5.5 \, {\rm fm}/c$\ ($3.7 \, {\rm fm}/c$) in the case of most 
central (peripheral) collisions, 
that happens in bulk (i.e. not surface 
dominated). This resembles the decoupling process of photons 
in the Early Universe. 
\par
It can be supposed that pion decoupling is caused mostly by a rapid 
cooling and dilution of the baryon density. 
In thermal models~\cite{Beccatini} the {\em chemical} freeze-out 
temperature ($\approx 170 \, {\rm MeV}$), needed to describe 
the observed particle ratios, is considerably higher than the 
{\em thermal} freeze-out temperature. 
This means that the relative abundances of particle species  
stop changing earlier than the end of (strong) elastic 
collisions (thermal freeze-out). The last stage of expansion would 
thus be characterized by  a volume dilution of baryon density 
without any appreciable particle recombination.
\section{Conclusions}
\label{sec_last}
The study of two-particle correlation functions 
has provided valuable 
information both on the geometry and the dynamical state of the 
reaction zone at freeze-out. 
A Coulomb correction procedure that assumes the same model 
for pion emission as that used in the HBT analysis has been 
implemented. 
\par
We have measured both negative and positive Yano-Koonin velocities 
in the LCMS system at slightly forward rapidity 
(with respect to $y_{cm}$), observing a strong increase of 
$v_{yk}$\ with the transverse pair momentum $K_t$.
\par
The simultaneous analysis of single-particle 
transverse mass spectra and two-particle correlation, with the help 
of a model that accounts for both longitudinal and transverse expansion 
of the source,  
has allowed for a reconstruction of the final freeze-out state of the 
source. We have measured the following parameters as a function 
of centrality: 
\begin{itemize}
\item the thermal freeze-out temperature, 
\item the rms transverse Gaussian radius, 
\item the average freeze-out proper time, 
\item the mean proper time duration of particle emission, 
\item the transverse flow velocity. 
\end{itemize}
From these data we have shown that our less central collision class (I)
exhibits a different dynamics: the indications are that the whole 
expansion develops less intensely (both longitudinally and in transverse 
direction), the duration of expansion is shorter, 
the final (i.e. at freeze-out) source dimension is smaller, 
the system ends up with a higher temperature  
than in the case of more central collisions. 
The source parameters of the two most central 
collision classes (III and IV) have been found to be very similar,  
thus suggesting the same geometrical and dynamical freeze-out state 
for these classes.  
Class II presents parameters that are intermediate between class I 
and class III-IV, its dynamics being however much closer 
to that of classes III--IV. 
\ack
We are grateful to U.~Heinz, B.~Tom\'{a}\v{s}ik and U.~A.~Wiedemann 
for fruitful discussion. We thank the NA49 and WA98 
Collaborations for having provided the numerical values of 
their HBT radii.
\end{document}